\documentclass{jfm}

\usepackage{relsize}
\usepackage{caption}
\usepackage{subcaption}
\usepackage{graphicx}
\usepackage{natbib}
\usepackage{mathrsfs} 
\usepackage{amsmath,mathtools}
\usepackage[usenames,dvipsnames]{color} 
\usepackage{ifthen}
\usepackage{tabularx}
\usepackage{longtable}
\usepackage{listings}
\usepackage{booktabs}
\usepackage{float}
\usepackage{makecell}
\usepackage{epstopdf}
\usepackage{pifont}

\newboolean{showcomments}
\setboolean{showcomments}{true}

\newcommand{\dennice}[1]{  \ifthenelse{\boolean{showcomments}}
{\textcolor{red}{(Dennice says:  #1)}}{}}

\ifCUPmtlplainloaded \else
  \checkfont{eurm10}
  \iffontfound
    \IfFileExists{upmath.sty}
      {\typeout{^^JFound AMS Euler Roman fonts on the system,
                   using the 'upmath' package.^^J}%
       \usepackage{upmath}}
      {\typeout{^^JFound AMS Euler Roman fonts on the system, but you
                   dont seem to have the}%
       \typeout{'upmath' package installed. JFM.cls can take advantage
                 of these fonts,^^Jif you use 'upmath' package.^^J}%
      }
  \else
  \fi
\fi

\ifCUPmtlplainloaded \else
  \checkfont{msam10}
  \iffontfound
    \IfFileExists{amssymb.sty}
      {\typeout{^^JFound AMS Symbol fonts on the system, using the
                'amssymb' package.^^J}%
       \usepackage{amssymb}%
         \let\leq=\leqslant
         
      }{}
  \fi
\fi

\ifCUPmtlplainloaded \else
  \IfFileExists{amsbsy.sty}
    {\typeout{^^JFound the 'amsbsy' package on the system, using it.^^J}%
     \usepackage{amsbsy}}
    {\providecommand\boldsymbol[1]{\mbox{\boldmath $##1$}}}
\fi

\providecommand\bcdot{\boldsymbol{\cdot}}

\newcommand\Real{\mbox{Re}} 
\newcommand\Imag{\mbox{Im}} 
\newcommand\Rey{\mbox{\textit{Re}}}  

\newsavebox{\astrutbox}
\sbox{\astrutbox}{\rule[-5pt]{0pt}{20pt}}

\newcommand\p{\ensuremath{\partial}}

\title[Convective velocity in turbulent channel flow]{An input-output based analysis of convective velocity in
	turbulent channels}

\author[Chang Liu, Dennice F. Gayme]%
{Chang Liu%
  \ns and Dennice F. Gayme \thanks{Email address for correspondence: dennice@jhu.edu}}

\affiliation{Department of Mechanical Engineering, Johns Hopkins University,
Baltimore, MD 21218, USA}

\pubyear{2010}
\volume{650}
\pagerange{119--126}

\date{?; revised ?; accepted ?. - To be entered by editorial office}

\newcommand{\bld}[1]{{\boldsymbol{ #1 }} } 
\DeclareMathAlphabet{\mathscrbf}{OMS}{mdugm}{b}{n} 
\newcommand{\e}[1]{{\text{e}^{ #1 }} } 
\newcommand{\dd}[2]{{ \frac{\text{d} {#1}}{\text{d} {#2}}   }} 
 
\newcommand{\im}{\text{i}}

\def\mline{\vrule width4pt height2.5pt depth -2pt}

\def\bdot{\raise.2em\hbox to .15em{.}}
\def\dashed{\mline\hskip3.5pt\mline\thinspace}
\def\dashdot{\mline\ \bdot\ \mline\thinspace}

\definecolor{gray}{gray}{0.5}

\def\bdotblack{\raise.25em\hbox to .15em{.}}

\newcommand{\cliu}[1]{{\color{black}#1}} 
\newcommand{\cliua}[1]{{\color{black}#1}} 

\newcommand{\cliub}[1]{{\color{black}#1}} 

\newcommand{\cliuc}[1]{{\color{black}#1}} 

\newcommand{\cliud}[1]{{\color{black}#1}} 

\newcommand{\cliue}[1]{{\color{black}#1}} 

\newcommand{\cliuf}[1]{{\color{black}#1}}

\newcommand{\rev}[1]{{\color{black}#1}}

\newcommand\cites[1]{\citeauthor{#1}'s\ (\citeyear{#1})}

\begin{document}
\large
\maketitle

\begin{abstract}
\large
This paper employs an input-output based approach to analyze convective velocities and the transport of fluctuations in turbulent channel flows. The convective velocity for a fluctuating quantity associated with streamwise--spanwise wavelength pairs at each wall-normal location is obtained through the maximization of the power spectral density associated with the linearized Navier-Stokes equations with a turbulent mean profile and delta-correlated Gaussian forcing. We first demonstrate that the mean convective velocities computed in this manner agree well with those reported previously in the literature. We then exploit the analytical framework to probe the underlying mechanisms contributing to the local convective velocity at different wall-normal locations by isolating the contributions of each streamwise--spanwise wavelength pair (flow scale). The resulting analysis suggests that the behavior \cliua{of the convective velocity in the near-wall region is} \cliue{influenced by large-scale} structures further away from the wall. \cliue{These structures} resemble Townsend's attached eddies in the cross--plane, yet show incomplete similarity in the streamwise direction. \cliub{We then} \cliud{investigate the role of each linear term in} \cliua{the momentum equation to isolate the} contribution of the pressure, mean shear, and viscous \cliud{effects} to the deviation of \cliud{the} convective velocity from the mean \cliua{at each flow scale.} \cliub{Our analysis highlights the role of the viscous effects, \cliue{particularly in regards to} large channel spanning structures \cliue{whose influence extends to the} near-wall region. \cliud{The results of this work suggest the promise of an input-output approach \cliue{for} \cliuf{analyzing} convective velocity \cliue{across} a range of flow scales \cliue{using only the mean velocity profile.} }}

\end{abstract}


\section{Introduction} 
\label{sec:intro}
Taylor's frozen turbulence hypothesis \citep{Taylor1938} and its variants have proven invaluable in the study of high Reynolds number wall-bounded turbulent
flows \citep{Marusic2010,Smits2011,LeHew2011}. However, its underlying assumption that the motion of turbulent fluctuations can be modelled as passive advection by the local mean velocity is known to break down in certain flow regimes \citep{Lin1953,Dennis2008,squire2017applicability}. In order to compensate for these known errors, \cliu{the local mean velocity is often replaced by a} convective velocity that better represents the spatio-temporal development of the fluctuations \citep{Zaman1981,Hutchins2011}. This convective velocity can be computed from simulation data \citep{Kim1993,delAlamo2009,chung2010large,lozano2014time,Geng2015,Renard2015} \cliu{\cliu{or} obtained from} \cliue{spatio-temporally resolved} experimental measurements \citep{Krogstad1998,lehew2010study,LeHew2011,deKat2015}. However, questions remain regarding \cliub{how to obtain an accurate estimate of} this quantity, \cliu{particularly in situations where the relevant data \cliuf{are} unavailable; e.g., in experiments using hotwire measurements or planar PIV.} In addition, there is not yet a full understanding of the mechanisms contributing to the convective velocity in each region of the flow. Such knowledge is required both to characterize the transport properties of fluctuating quantities, and to identify when direct application of Taylor's hypothesis with the mean velocity is insufficient.

An early work by \citet{Lin1953} suggests that Taylor's hypothesis works well when the mean flow is approximately spatially uniform and when turbulence intensities are low, \cliu{but} breaks down in region\cliu{s} of high shear. \citet{Lumley1965} further refines \cliu{this} spatial uniformity requirement, \cliu{suggesting} that weak interaction\cliu{s} between eddies of different sizes are also \cliud{needed} \cliu{to ensure} the validity of Taylor's hypothesis. \citet{Geng2015} \cliub{provide support for the break down of Taylor's hypothesis in highly sheared \cliud{regions of the flow} by} \cliua{explicitly computing the contribution of advection by the mean flow (Taylor's hypothesis) to convective velocity in the viscous sublayer using \cliuf{Direct Numerical Simulation (DNS)} data from channel \cliud{flows} at $Re_\tau=205$ and $932$}. \cliua{In particular, they compute} the average amplitudes of different terms in the momentum equation through DNS and illustrate that advection by the mean flow \cliud{provides} less than 50\% of the streamwise momentum \cliud{flux} \cliu{in the viscous sublayer}. Taylor's hypothesis has also \cliub{proven to inadequately describe the convection of} large-scale components of the flow. \citet{Dennis2008} compare the spatial evolution of a turbulent flow inferred from the temporal information using Taylor's hypothesis with those obtained using Particle Image Velocimetry (PIV) at a wall parallel plane sufficiently removed from the wall so that the assumptions of \citet{Lin1953} are satisfied. The authors find that even though the PIV fields are \cliud{qualitatively} similar,  several large-scale features are not reproduced in the fields generated using Taylor's hypothesis. 

\cliua{The \cliud{validity of Taylor's hypothesis across a range of flow scales is}} explored by \citet{Fisher1964} \cliua{who use} two-point space-time correlation for statistically stationary turbulence to compute the convective velocity of streamwise velocity fluctuations as a function of streamwise
spatial and temporal separation; i.e., a streamwise (or temporal) scale-dependent convective velocity. \citet{Fisher1964} and subsequently \citet{favre1967structure,Zaman1981} show that the convective velocity computed in this manner does not coincide with the local mean velocity and can be strongly dependent on the \cliuf{streamwise spatial or temporal separation}. 
A similar phenomenon has also been observed in measurements of wall-pressure \citep{Willmarth1962}.

\cliud{This scale dependence, particularly the increasing deviation of the convective velocity from the mean flow as spatial separation is increased, was}  identified as a possible explanation for \cliub{the known} \cliu{discrepancy between the convective and mean velocities \cliua{near} the wall}. In particular, the larger convective velocity \cliub{versus the local mean velocity in the near-wall region} has been attributed to faster moving structures centered further away from the wall whose influence extends to the wall due to their large size \citep{Dinkelacker1977,Kreplin1979,Farabee1991,Kim1993,Hutchins2011}. \citet{delAlamo2009} find that fast and wide streamwise elongated structures, coherent up to the core region, provide a consistent contribution to the energy-weighted average convective velocity close to the wall. They relate these structures to the large modes \citep{Bullock1978,delAlamo2004} \cliua{reminiscent of} Townsend's `inactive' eddies \citep{Townsend1961,Bradshaw1967} \cliub{and the very large-scale motions \citep{Guala2006,Balakumar2007,Hutchins2007,monty2007large}, which \cliud{have been shown to modulate} small-scale structures \citep{mathis2009large,mathis2009comparison,ganapathisubramani2012amplitude,yang2018implication}}. \cliu{Although these works \cliud{provide evidence that} scale interactions \cliud{contribute to} the breakdown of Taylor's hypothesis at the wall, a full understanding of the \cliua{underlying} mechanisms \cliud{across the full range of flow regimes} has yet to be realized.}

\rev{In this paper, we \cliua{explore \cliub{the} mechanisms \cliub{underlying the convective velocity \cliue{of fluctuating quantities} in wall-bounded turbulence} using} a spatio-temporal transfer function that enables us to isolate the contributions and interactions across the full range of flow scales. This approach \cliu{allows us to \cliua{compute quantities for a range of Reynolds numbers given \cliue{an associated turbulent} mean velocity profile.}} Our \cliub{analytical framework} is based on stochastically-forced \cliua{Linearized} Navier-Stokes (LNS) equations, which have a long history in the study of wall-bounded shear flows; e.g., in characterizing energy amplification associated with stochastic disturbances \citep{Farrell1993,Bamieh2001} and isolating the most sensitive input-output paths \citep{Jovanovic2005}.} \rev{The LNS equations have also proven useful in characterizing coherent structures \citep{Smits2011,mckeon2017engine,jimenez2018coherent}. For example, low rank approximations of the resolvent operator \citep{McKeon2010}  have been used to explain the scalings of the very large-scale motions and to reconstruct the packet hairpin vortices \citep{Sharma2013}. \citet{Luhar2014} \cliud{also} use the resolvent framework to predict the high-amplitude wall-pressure previously observed in experiments and simulations.  \citet{Moarref2013} combine this framework with term balancing arguments to reproduce the inner, outer, and geometrically self-similar scalings of the streamwise energy density in turbulent channel flows. \cliuf{Input-output analysis of the NS equations linearized about a base profile with an eddy viscosity term \citep{reynolds1972mechanics} leads to accurate predictions of the spanwise spacing of near-wall streaks \citep{delAlamo2006,cossu2009optimal,pujals2009note,hwang2010amplification,Hwang2010Linear} and large-scale structures \citep{illingworth2018estimating,madhusudanan2019coherent,morra2019relevance}.} \cliud{Related work employing the impulse response \citep{vadarevu2019coherent} of the LNS transfer function has led to self-similar vortex structures.}

} 

\cliud{The transfer function of the LNS has \cliue{previously} been used to compute quantities associated with the convective velocity of a fluctuating quantity in wall-bounded turbulent flows. For example, the resolvent framework was used to show that the phase speed of streamwise velocity fluctuations with peak contribution to the energy density deviates from the mean velocity in the near-wall region \citep{Moarref2013}. \citet{Luhar2014} also used the resolvent framework to investigate the scale dependence of wall-pressure propagation speed, which showed agreement with \cliuf{the} empirically determined convective velocity \citep{panton1974wall}. \citet{Zare2017} also computed \cliub{the} convective velocity of streamwise velocity fluctuations for \cliue{one} specific flow scale} \cliud{based on the LNS equation} with temporally correlated (colored) stochastic forcing. Their results show qualitatively similar behavior to convective velocities obtained by \citet{delAlamo2009}. These works demonstrate the utility of transfer function based \cliuf{approaches} in computing \cliud{the} convective velocity. \cliub{However, none of these works employed input-output analysis to  investigate the underlying mechanisms that lead to the deviation of convective velocity from the local mean velocity.}

\rev{This work takes steps in that direction by using an input-output framework to systematically investigate the scale-dependent convective velocity of velocity fluctuations in turbulent channels. We begin by demonstrating that the \cliuf{proposed} approach provides good agreement with the mean convective velocity predictions \cliua{for fluctuations of the three velocity components} previously published in the literature \citep{Kim1993,Geng2015}. We then \cliua{direct our attention to the streamwise velocity fluctuations and} exploit the analytical framework to compute \cliud{the convective velocity for each streamwise--spanwise wavenumber pair at different wall normal locations} and \cliue{to} \cliu{examine interactions} between different scales. \cliu{\cliua{The results of our scale-dependent analysis} are consistent with those obtained using DNS data \citep{delAlamo2009}. In particular,} \cliua{the} convective velocities predicted using the \cliuf{input-output based approach employed here} show more variation with scale closer to the wall, \cliu{with the largest variation occurring in the viscous sublayer.} Our analysis suggests that \cliub{this} viscous sublayer behavior arises due to structures that \cliub{are self-similar in the spanwise and wall-normal plane and} scale with wall-normal height, which supports Townsend's attached eddy hypothesis regarding the dominant dynamical structures in wall-bounded flows.

Finally, \cliu{in the spirit of \cites{Lin1953} term--by--term analysis of the momentum equation,} we examine how each \cliud{linear} term in the \cliud{momentum equation} contributes to the deviation of the convective velocity from the mean. A linear analysis is expected to yield insight in this regard because both the mean shear term and the viscous term, which play critical roles \cliu{in the dynamics of the} near-wall region, are linear. Moreover, \cliue{it was recently shown that} resolvent analysis retains the fast pressure component arising from the linear interaction between the mean shear and turbulent wall-normal velocity \citep{Luhar2014}, \cliua{and therefore our computations also include these phenomena.} \cliub{\cliud{Our analysis employs an expression for} scale-dependent convective velocities \cliud{derived by \citet{delAlamo2009}, who did not further analyze \cliue{the various terms}. This work also builds upon that of} \citet{Geng2015} who quantify each term's contribution to \cliud{the} convective velocity at different wall-normal locations.} \cliub{Our \cliue{results indicate} that} \cliua{the viscous term \cliub{provides the largest contribution to the} deviation of the convective velocity from the mean in the near-wall region.} \cliub{Based on these observations, we propose} a viscous correction to Taylor's hypothesis, \cliub{and demonstrate that the revised model accurately} \cliu{reproduces the behavior of the near-wall convective velocity for large--scale structures.}
}

The remainder of the paper is organized as follows. Section \ref{sec:problem} describes the problem setup. We \cliue{detail} our \cliue{transfer function} based approach and numerical scheme for calculating the convective velocities in sections \ref{sec:method} and \ref{sec:numerical}, respectively. \cliua{In section \ref{sec:convective}, we apply the \cliuf{input-output based approach}} using mean velocity profiles from turbulent channel flows obtained from \cliu{\citet{Lee2015}} \cliud{at three different Reynolds numbers ($Re_\tau=550,\;1000,$ and $5200$)}. \cliua{We then discuss} the physical origin of the near-wall convective velocities. Section \ref{sec:coher} explores the wall-normal coherence of the structures contributing to the convective velocities at a particular wall-normal location. Section \ref{sec:term} analyzes the contribution of each of the linear terms \cliud{in the momentum equation} \cliu{to the total convective velocity for each streamwise--spanwise wavenumber pair (flow scale)}. Based on this term-by-term analysis, a viscous correction to Taylor's hypothesis is proposed and discussed. Section \ref{sec:conclusion} concludes the paper.

\section{Problem setup}
\label{sec:problem}
\begin{figure}
	(a) \hspace{2.85in} (b) 
	
\centering
\includegraphics[scale=0.48]{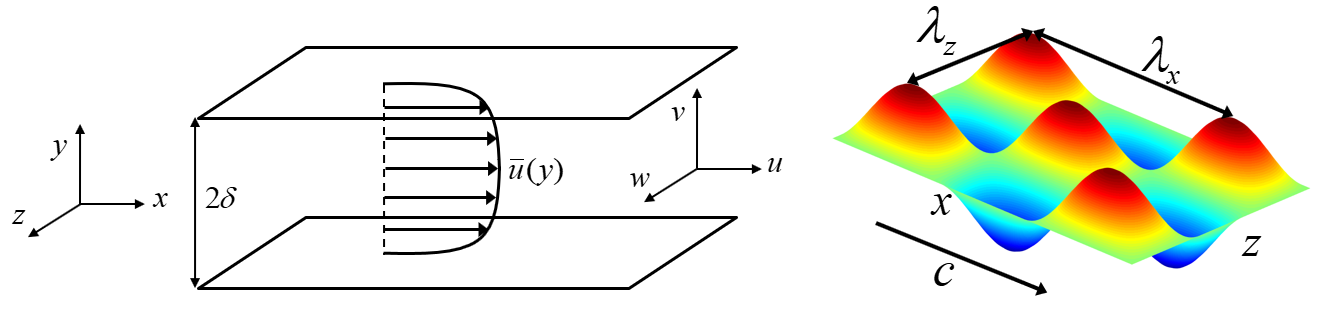}
\caption{(a) Turbulent flow between two infinite parallel plates with mean profile $\bar{u}(y)$. (b) The fluctuations $\bld{u}'$ are decomposed into \cliuf{traveling} waves with wavelengths $\lambda_x$, $\lambda_z$ in the $x$, $z$ directions and downstream phase speed $c=-\lambda_x\omega/2\pi$ using equation (\ref{FTdefn}).}
\label{fig:schematic}
\end{figure}

We consider incompressible flow between two infinite parallel plates driven by a streamwise pressure gradient as shown in figure \ref{fig:schematic}(a), where \cliu{$x,y,z$ are the streamwise, wall-normal, and spanwise directions, respectively.} We decompose the velocity field, $\bld{u} = \begin{bmatrix} u & v & w\end{bmatrix}^{\mathsf{T}}$, \cliub{with $^{\mathsf{T}}$ indicating the transpose}, and the pressure field, $p$, into mean and fluctuating quantities; i.e.,  $\bld{u} = \bar{u}(y)\boldsymbol{i} + \bld{u}'$ \cliud{with $\boldsymbol{i}$ denoting the streamwise unit vector} and $p = \bar{p} + p'$, where the overbars indicate time averaged quantities, $\bar{\phi} = \lim_{T\rightarrow\infty} \frac{1}{T}\int_{0}^{T} \phi(t)\,dt$, and primes indicate fluctuating quantities. The \cliub{dynamics of the} fluctuations $\bld{u}'$ and $p'$ are governed by
\begin{subequations} \label{NS_All}
\begin{align}
\p_{t} \bld{u}'   
+  \bar{u}\p_x \bld{u}'  +\nabla p'+ v'  \dd{\bar{u} }{y}\boldsymbol{i} 
-\frac{1}{\Rey_\tau}\Delta \bld{u}'
 &=\cliud{- \bld{u}' \bcdot \nabla \bld{u}' + \overline{\bld{u}' \bcdot \nabla \bld{u}'}}, \label{NSDecompf1} \\
\nabla \bcdot \bld{u}'=0. \label{NSDecompf2}
\end{align}
\end{subequations} 
The spatial variables are normalized by \cliu{the} half channel height $\delta$; e.g., $y=y_*/\delta \in [-1,1]$, where the \cliu{subscript $*$} indicates dimensional quantities. The velocity is normalized by the friction velocity \cliu{$U_{\tau} = \sqrt{\tau_{\text{w}}/\rho }$, \cliub{where} $\tau_{\text{w}}$ is the time-averaged mean shear stress at the wall\cliua{,} and $\rho$ is the density of the fluid, \cliub{which leads to} $u:=u^+=u_*/U_{\tau}$.\footnote{Note, in (\ref{NS_All})\cliua{,} we omit the $+$ \cliud{superscripts for the velocity fluctuations} for notational convenience.}} Time and pressure are normalized by $\delta/U_\tau$, and $\rho U^2_\tau$, respectively. We define the inner unit length scale \cliub{as} $\delta_v=\nu/U_\tau$ and use the superscript $+$ to denote the distances measured in inner units; i.e., $y^+=y_*/\delta_v$. The friction Reynolds number is defined as $\Rey_{\tau} =\delta U_{\tau}/\nu=\delta^+$.

\cliuf{Invariance to shifts in $(x,z,t)$ of equation\cliua{s} (\ref{NS_All}) allows us to \cliub{employ} the $(x,z,t)$ spatio-temporal Fourier transform\cliub{,}}
\begin{align}
\hat{\psi}(y;k_x,k_z,c) = \mathcal{F}(\psi) \equiv \int\limits_{-\infty}^{\infty}\int\limits_{-\infty}^{\infty}\int\limits_{-\infty}^{\infty}\psi(x,y,z,t)\e{-\im(k_x ( x  - ct) + k_z z )}\,\text{d}t\,\text{d}x\,\text{d}z,\label{FTdefn}
\end{align} 
where $k_x = 2\pi/\lambda_x$ and $k_z = 2\pi/\lambda_z$ are the respective dimensionless $x$ and $z$ wavenumbers normalized by the channel half-height $\delta$. \cliub{The transform (\ref{FTdefn}) decomposes} the flow into traveling waves with wavelengths $\lambda_x$, $\lambda_z$ in the $x$, $z$ directions and downstream phase speed $c=-\omega/k_x=-\lambda_x\omega/2\pi$, where $\omega$ is the frequency; \cliu{see figure \ref{fig:schematic}(b).} Applying the Fourier transform to (\ref{NS_All}) and denoting the nonlinear term as: \cliud{$\bld{f}'(x,y,z,t) \equiv  - \bld{u}' \bcdot \nabla \bld{u}' + \overline{\bld{u}' \bcdot \nabla \bld{u}'}$} yields
\begin{subequations}\label{NSPFT_all}
\begin{align}   
\underbrace{\im k_x (\bar{u} - c)\bld{\hat{u}}' }_{\text{I}}
\underbrace{+ \hat{\nabla} \hat{p}' }_{\text{IIa}} \underbrace{+\hat{v}'  \dd{\bar{u} }{y} \bld{i}}_{\text{IIb}} \underbrace{-\frac{1}{\Rey_\tau}\hat{\Delta} \bld{\hat{u}}'}_{\text{IIc}} 
    &=\underbrace{ \bld{\hat{f}}'  }_{\text{III}} \label{NSPFT2}, \\
\hat{\nabla} \bcdot \bld{\hat{u}}'&=0, \label{NSPFT1} 
\end{align}
\end{subequations}
where $\hat{\nabla} =\begin{bmatrix} \im k_x & \partial_y & \im k_z \end{bmatrix}^{\mathsf{T}}$ \cliub{and} $\hat{\Delta} = \partial_{yy} - (k_x^2 + k_z^2)$. The corresponding no-slip boundary conditions are given by
\begin{align}   
\bld{\hat{u}}'(y=\pm1;k_x,k_z,c)=\begin{bmatrix} 0 & 0 & 0 \end{bmatrix}^{\mathsf{T}},\; \forall k_x, k_z,c \in \mathbb{R}. \label{NSBC}
\end{align}

The terms in (\ref{NSPFT2}) can be grouped into: (I) advection of the fluctuations by the mean velocity in the frame of reference of the \cliuf{traveling} wave fluctuations\cliua{,} \cliue{terms} \rev{(IIa)-(IIc), \cliu{which capture the} \cliud{respective effects of pressure, shear, and viscosity\cliua{,}} and} (III) fluctuation-fluctuation nonlinear interactions. 

Taylor's frozen turbulence hypothesis states that for sufficiently low turbulence intensities ($|\bld{u}'_{\text{rms}}|_{\infty}/\bar{u} \rightarrow 0$), the spatio-temporal development of turbulent fluctuations can be described as downstream advection by the mean velocity $\bar{u}(y)$ \citep{Taylor1938}. This statement is equivalent to setting all terms except (I) in (\ref{NSPFT2}) to zero, which reduces the equation (\ref{NSPFT2}) to the passive advection model\cliua{:}
\begin{align}
\im k_x (\bar{u}  - c) \bld{\hat{u}}' = 0 \label{Taylors1}.
\end{align}  
As previously discussed, \cliu{the direct application of Taylor's hypothesis} tends to be inaccurate in regions \cliu{near the wall}, where \cliub{the terms in} (IIa)-(IIc) and (III) in (\ref{NSPFT2}) provide non-negligible contribution\cliu{s}\cliub{; see e.g., \citet{Lin1953,Geng2015}}. \cliua{It is well-known that} \cliub{the} model in (\ref{Taylors1}) can be improved by replacing the mean velocity, $\bar{u}$, with an empirically determined \cliud{mean} convective velocity $\psi_c$ \citep{Zaman1981,Hutchins2011} for the fluctuating quantity of interest. In the next section, we describe \cliuf{an input-output} based approach to computing such a $\psi_c$.

\section{Method for calculating scale-dependent convective velocities}
\label{sec:method}
In this section, we describe \cliuf{the employed} input-output approach \cliu{to computing spatio-temporal convective \cliub{velocities of fluctuating quantities} given a mean velocity profile \cliud{$\bar{u}(y)$}}.
\cliub{First we rewrite} (\ref{NSPFT_all}) in the form
\begin{align}
\mathcal{L} 
\cliu{\begin{bmatrix} \bld{\hat{u}}'  \\\hat{p}' \end{bmatrix}} 
= \mathcal{B} \bld{\hat{f}}', \label{OpFormEqns}
\end{align}
where 
\begin{align}
\mathcal{L} &:=
\begin{bmatrix}
(\im k_x (\bar{u} - c)  
- \frac{1}{Re_\tau}\hat{\Delta})\mathbf{I}_{3\times 3} +   \dd{\bar{u} }{y} \mathbf{S} & \hat{\nabla} \\
\hat{\nabla}^\mathsf{T} & 0 
\end{bmatrix}, 
\qquad
\mathcal{B} :=
\begin{bmatrix}
\mathbf{I}_{3\times 3} \\
\mathbf{0}_{1\times 3}
\end{bmatrix}, 
\label{sysOpsDefn}
\end{align}
and \cliuf{$\bld{\hat{f}}'$} is parameterized as input forcing. \cliub{In equation (\ref{sysOpsDefn}), $\mathbf{I}_{n\times n}$ and $\mathbf{0}_{m\times n}$ are respective identity and zero matrices with their sizes indicated by their subscripts, and $\mathbf{S}:=\scriptsize\begin{bmatrix}0 & 1 & 0\\
0 & 0 & 0\\
0 & 0 & 0\end{bmatrix}$.} \cliuf{A non-bold symbol represents a scalar; e.g., the $0$ appearing in (\ref{sysOpsDefn}) is a scalar quantity.} We \cliub{then} define an output variable
\begin{align}
\hat{\psi}' =\mathcal{C}_{\hat{\psi}'}\cliu{\begin{bmatrix} \bld{\hat{u}}'  \\ \hat{p}' \end{bmatrix},}
\label{outputEqn}
\end{align}
where $\mathcal{C}_{\hat{\psi}'} = \mathcal{C}_{\hat{\psi}'}(y;k_x,k_z)$ is a linear operator \cliub{that maps the state variables to the} output of interest. Here, we use the primitive variables \cliu{$\bld{\hat{u}}'$ and $\hat{p}'$} as states rather than the \cliu{wall-normal velocity and vorticity coordinates of the} commonly studie\cliu{d} Orr-Sommerfeld and Squire equations because this form of \cliu{the} equations provides direct information about the pressure \citep{Luhar2014}, which we will later exploit in the term-by-term analysis in section \ref{sec:term}.  

\cliu{We} define the input-output map $\mathcal{G}_{\hat{\psi}'}$ between the input $\bld{\hat{f}}'$ and the output $\hat{\psi}'$, in \cliub{the manner of} \citet{McKeon2010,Luhar2014,mckeon2017engine} as
\begin{align}
\hat{\psi}' =\mathcal{C}_{\hat{\psi}' }\mathcal{L}^{-1} \mathcal{B} \bld{\hat{f}}' = \mathcal{G}_{\hat{\psi}'}(y;k_x,k_z,c)\bld{\hat{f}}'.
\label{IOmap}
\end{align}

\cliu{The convective velocity of a fluctuating variable $\psi^\prime$ can be computed using the following two-point space-time correlation for statistically stationary turbulence:
\begin{align}
R_{\hat{\psi}'}(\xi,\tau;\bld{x}) = \langle \psi'(\bld{x},t)\psi'(\bld{x}+\xi \boldsymbol{i},t + \tau) \rangle,\;\; \psi=u,v,w,\omega_x,\omega_y,\omega_z,p\cliua{,}
\label{twoptcorrDefn}
\end{align}
where $\xi$ and $\tau$ are the \cliua{respective} streamwise and temporal separation between two points. Convective velocities for fluctuations $\psi'$ at some $\bld{x}$ are computed from \eqref{twoptcorrDefn} by fixing either $\xi$ or $\tau$ and varying the other separation variable to maximize $R_{\hat{\psi}'}(\xi,\tau;\bld{x})$ \citep{Wills1964,Fisher1964,Kim1993,Zaman1981,Krogstad1998}.} \cliub{We adapt this idea to our \cliuf{approach} by computing} \cliu{the Power Spectral Density (PSD) \citep{Wills1964,delAlamo2009} \cliub{for} the input-output map (\ref{IOmap}) as}
\begin{align}
\Phi_{\hat{\psi}'}(y; k_x, k_z, c) = 
\langle\hat{\psi}'\hat{\psi}'^*\rangle
&
=   \mathcal{G}_{\hat{\psi}'} \langle\bld{\hat{f}}'  \bld{\hat{f}}'^*\rangle \mathcal{G}_{\hat{\psi}'}^* 
=   \mathcal{G}_{\hat{\psi}'}  \mathcal{G}_{\hat{\psi}'}^* \label{phipsi1}
\end{align}
with $\bld{{f}}'(x,y,z,t)$ parametrized as spatio-temporal delta-correlated Gaussian noise with unit variance; i.e., noise that is white in space and time \citep{Jovanovic2005}. The \cliu{superscript $*$} \cliua{in (\ref{phipsi1})} denotes the complex conjugate, and the angle brackets $\langle\;\rangle$ indicate an ensemble \cliub{averaging operation}. 

\cliu{The convective velocity $\psi_c$ is then obtained as}
\begin{align}
\cliu{\psi}_{c}(y;k_x,k_z) \equiv \text{arg max}_{c} \Phi_{\hat{\psi}'}(y; k_x, k_z, c), \label{ucdefn}
\end{align}
which represents convective velocities of the coherent structure\cliu{s} with \cliu{$(x,z)$ spatial extents of} $\lambda_x=2\pi/k_x$ and $\lambda_z=2\pi/k_z$ \cliu{as a function of wall-normal location}. \cliuf{This definition of convective velocity based on (\ref{ucdefn}) neglects the distribution of the PSD for a given ($\lambda_x, \lambda_z$) pair, which is expected to contain energy at a range of temporal frequencies. The distribution of the spectrum could be partially accounted for by instead defining the convective velocity in terms of the center of gravity of the PSD. That quantity is commonly used to compute convective velocity in simulation (DNS and LES) studies as it requires time-averaging instead of Fourier transforming in the time domain; see e.g., \citet{delAlamo2009,chung2010large,Renard2015}. Our approach can be adapted to accommodate such a definition (and others) through a suitable modification of equation (3.7). In the current work, we recomputed a subset of the results using the center of gravity method to ensure that the main conclusions of our study are not altered by our choice of definition.}

{\color{blue}
\rev{\cliu{\cliua{Assuming that $\bld{{f}}'(x,y,z,t)$ is spatio-temporal delta-correlated Gaussian noise with unit variance} implies that the velocity itself is Gaussian\cliub{. This} is clearly not true \cliub{as} velocity probability density functions are known to have heavy tails and odd-order moments \cliua{that} do not vanish \citep{Frisch1995}\cliu{. Colored-in-time forcing has  been shown to produce more accurate statistics} \citep{Zare2017}. However, the Gaussian white-noise parametrization is appealing because it is a simple, analytically tractable forcing that has been widely used to provide important insights into the dynamics; e.g., \citet{Farrell1993,Bamieh2001,Jovanovic2005}. \rev{Therefore, this} type of forcing provides a good starting point for understanding the role of linear mechanisms in determining the convective velocity and \cliua{simplifies analysis} because it does not introduce a preferential forcing in any of the spatial or temporal directions. 
}}

}

\cliu{\cliua{We focus on} streamwise, wall-normal, and spanwise velocity fluctuations which are computed based on the respective output operators}
\begin{align}
\mathcal{C}_{\hat{u}'} = \begin{bmatrix}
 1 & 0 & 0 & 0
\end{bmatrix},\quad
\mathcal{C}_{\hat{v}'} = \begin{bmatrix}
  0 & 1 & 0 & 0
\end{bmatrix}, \quad \mbox{ and }\quad
\mathcal{C}_{\hat{w}'} = \begin{bmatrix}
 0 & 0 & 1 & 0
\end{bmatrix}
\label{output_velocity_Coeffs}
\end{align} 
in (\ref{IOmap}). \cliua{However, we note that the approach} can be generalized to the calculation of the convective velocity for any \cliu{fluctuating quantity}, $\psi'$, whose Fourier transform can be written in the form (\ref{outputEqn}) with an appropriate choice of $\mathcal{C}_{\hat{\psi}'}$. For example, \cliu{the output operators corresponding to} the vorticity fluctuations $\hat{\omega}_x'$, $\hat{\omega}_y'$, $\hat{\omega}_z'$, and the pressure $\hat{p}'$ \cliu{are given by}:
\begin{align}
&\mathcal{C}_{\hat{\omega}_x'} = \begin{bmatrix}
0 & -\im k_z & \p_y & 0 
\end{bmatrix}, \quad
\mathcal{C}_{\hat{\omega}_y'} = \begin{bmatrix}
\im k_z & 0 & -\im k_x & 0
\end{bmatrix}, \nonumber\\
&\mathcal{C}_{\hat{\omega}_z'} = \begin{bmatrix}
-\p_y & \im k_x & 0 & 0
\end{bmatrix}, \quad
\mathcal{C}_{\hat{p}'} = 
\begin{bmatrix}
0 & 0 & 0 & 1
\end{bmatrix}\cliua{,}
\label{output_vorticity_pressure_Coeffs}
\end{align} 
respectively. An analysis of the convective velocity of vorticity fluctuations is carried out in \citet{liu2019vorticity}.

In the next section, we describe \cliu{the numerical implementation of the \cliuf{input-output based approach} for channel flow at three different Reynolds numbers. The resulting convective velocities are analyzed in subsequent sections.}

\section{Numerical approach}
\label{sec:numerical}

The operators in (\ref{phipsi1}) are discretized using the Chebyshev differentiation matrices generated by the {\sc{Matlab}} routines of \citet{Weideman2000}. \rev{\cliub{We denote the} discretization of $\mathcal{G}_{\hat{\psi}'}$ \cliub{as $\mathcal{\widetilde{G}}_{\hat{\psi}'}$. The resulting discretized} expression for the \rev{PSD} at wall-normal location $y = y_i$ \cliub{is given by}
\begin{align}
 \widetilde{\Phi}_{\hat{\psi}'}(y_i; k_x, k_z, c)\!=\!(\widetilde{\Phi}_{\hat{\psi}'}(\mathbf{y}; k_x, k_z, c) )_{i}\!=\!\left(\mathcal{\widetilde{G}}_{\hat{\psi}'}(\mathbf{y}; k_x, k_z, c)\mathcal{\widetilde{G}}_{\hat{\psi}'}^* (\mathbf{y}; k_x, k_z, c) \right)_{ii},  \label{phipsiFD}
\end{align}
where $\mathbf{y} = \{ y_1,y_2,... \}$ are the discrete grid points in the wall-normal direction\cliua{,} and $(\mathsfbi{A})_{ij}$ indicates the $(i,j)$ element of the matrix $\mathsfbi{A}$.} The convective velocity at a fixed $(y;k_x,k_z)$ can then be approximated \cliub{through the discretized analog of} (\ref{ucdefn}).
\cliue{In computing this quantity, we employ the} Clenshaw--Curtis quadrature \citep{trefethen2000spectral} to obtain \cliu{the} $L_2$ inner product for \cliub{the Chebyshev} spaced wall-normal grid. We implement the no slip boundary condition $\bld{\hat{u}}'(y=\pm 1)=\bold{0}$ explicitly \cliu{following the approach of} \citet{Luhar2014}. This implementation allows us to \cliud{use} primitive variables $ \bld{\hat{u}}'$ and $\hat{p}'$, which \cliub{as previously discussed} offers us direct information \cliue{regarding the fast} pressure. \cliua{We performed the same analysis using the Orr-Sommerfeld and Squire form described in \citet{Jovanovic2005} and verified that results do not change.} 

The turbulent mean velocities \cliub{in (\ref{NSDecompf1})} are obtained from the DNS of \citet{Lee2015} \cliua{at $Re_\tau=550,\;1000,$ and $5200$}, \cliuf{which all use simulation domains with $L_{x*}/\delta=8\pi$ and $L_{z*}/\delta=3\pi$.} For \cliua{the} $Re_\tau=550$ and $Re_\tau=1000$ cases, our calculations use 122 collocation points in the wall-normal \cliuf{direction}. \cliua{We employ} 192 collocation points for the $Re_{\tau}=5200$ calculations. \cliuf{We compute the optimal value of equation (\ref{ucdefn}) by computing the PSD for 201 uniformly spaced points over the phase speed range $c^+ \in [0,30]$ for each wavenumber pair $(k_x,k_z)$. We then select the single $c^+$ that maximizes the PSD.} \cliuf{This phase speed range and }$90 \times 90$ logarithmically spaced points in the spectral range \cliub{$k_x \in [10^{-2},10^3]$ and $k_z \in [10^{-2},10^3]$} \cliuf{are employed} for all three Reynolds numbers. We verified that \cliuf{doubling either} the number of collocation points in the wall-normal direction or \cliuf{the number of Fourier modes in the horizontal directions} does not alter the results, indicating grid convergence.

\section{Convective velocity in turbulent channels}
\label{sec:convective}

\rev{In this section, we use the method described in sections \ref{sec:method} and \ref{sec:numerical} to compute \cliu{the convective velocity} of the velocity fluctuations \cliub{for} turbulent channel flow at $\Rey_{\tau}=550$, $\Rey_{\tau}=1000$, and  $\Rey_{\tau}=5200$.} We first validate the approach by computing the mean convective velocities and comparing our results to those \cliub{computed from DNS data \citep{delAlamo2009,Geng2015}}. \cliub{We then} take advantage of the analytical framework to further analyze the contribution of different length scales to the local convective velocity. 

\subsection{Validation of the input-output based approach}
\label{ssec:uc1D}

The weighted average convective velocity $[\cliu{\psi}_c]_{h}$ of \cliub{a} fluctuating quantity $\psi'$ can be computed as\cliua{:}

\begin{align}
[\cliu{\psi}_c]_{h}(y)
&= 
\frac{ \int_{\Omega} \cliu{\psi}_c(y;k_x,k_z) h(y;k_x,k_z)\,\text{d}k_x \text{d}k_z}{ \int_{\Omega}  h(y;k_x,k_z) \,\text{d}k_x \text{d}k_z} \label{uc1Ddefn}
\end{align}
with an averaging domain $\Omega$ over $(k_x,k_z)$ and a weighting function $h(y;k_x,k_z)=\langle|\mathcal{F}_{xz}(\psi')|^2\rangle k_x^2$\rev{, where $\mathcal{F}_{xz}$ is the $x$-$z$ Fourier transform:
\begin{align}
\mathcal{F}_{xz}(\psi')(y;k_x,k_z,t) \equiv \int\limits_{-\infty}^{\infty}\int\limits_{-\infty}^{\infty}\psi'(x,y,z,t)\e{-\im(k_x x + k_z z )}\,\text{d}x\,\text{d}z. \label{spatialFTDefn}
\end{align}

\cliub{We compute this average quantity for each of the three fluctuating velocity components by first computing the convective velocity of each component using} \cliu{(\ref{phipsi1}) and (\ref{ucdefn}) with the \cliub{corresponding} output operators given in (\ref{output_velocity_Coeffs}). \cliub{These quantities are then filled into (\ref{uc1Ddefn}) with an averaging domain $\Omega:(\lambda_x^+,\lambda_z^+)>(500,80)$.} \cliub{The weighting function $h(y;k_x,k_z)$ in equation (\ref{uc1Ddefn}) is selected to provide the least-squares fit to the passive advection model: $\p_t \psi' + [\cliu{\psi}_c]_h \p_x \psi' = 0,\;\psi'=u',v',w'$ as discussed by \citet{delAlamo2009}. This choice allows a direct comparison with \citet{Geng2015}, who explicitly employed this fit in their computations.}
} 

}

\begin{figure}
    \centering
    \includegraphics[width=3.3in]{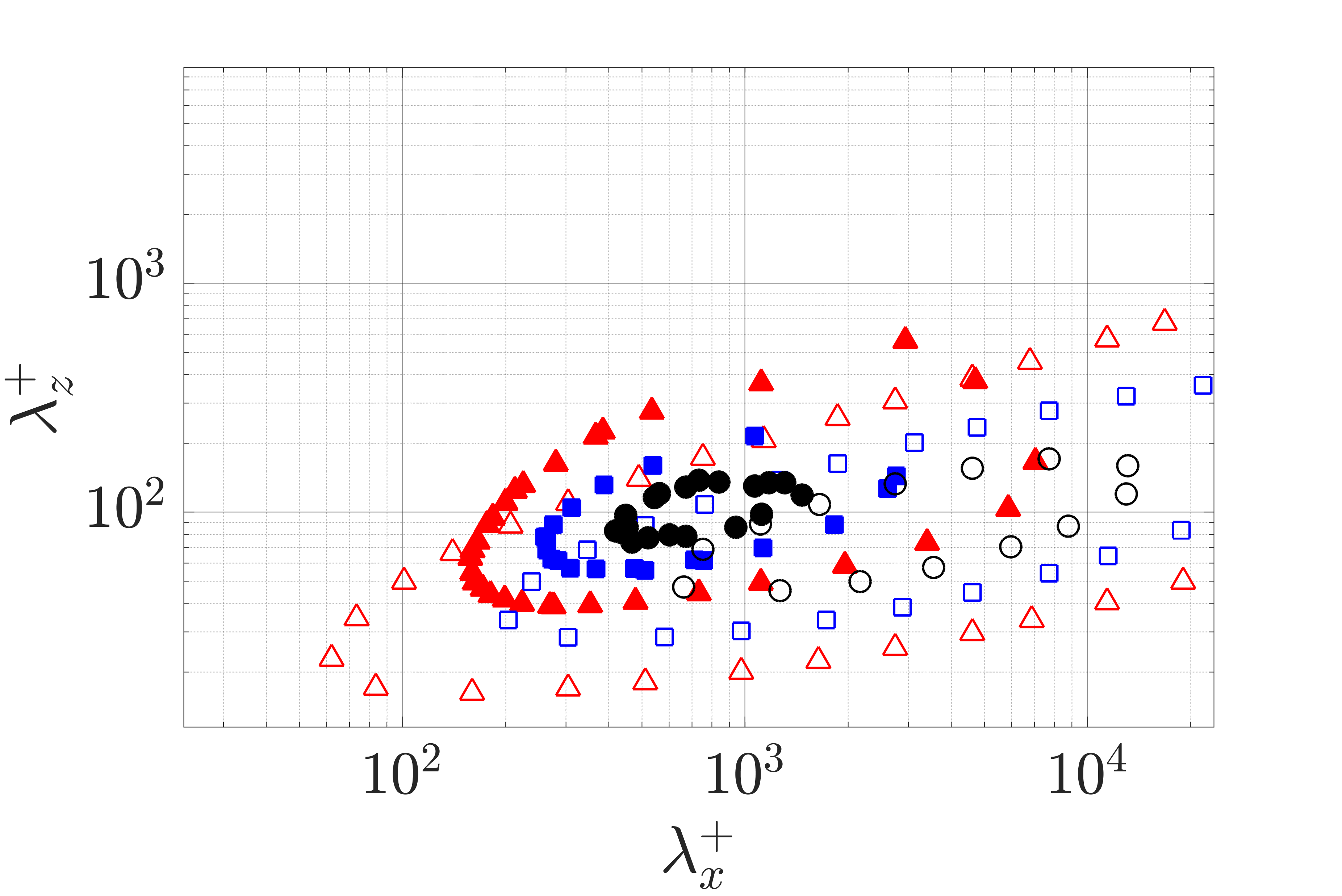}
    \caption{\cliuf{Premultiplied two-dimensional spectral density of streamwise velocity fluctuations $k_xk_z\int \Phi_{\hat{u}'}(y;k_x,k_z,c)dc$ at $y^+\approx 5$ for Reynolds number $Re_\tau=1000$. Contours are shown for 0.2 ({\color{red}\parbox{0.0825in}{$\hspace{-0.0070in}\vspace{-0.015in}\mathlarger{\triangle}$}}$\,$); 0.5 ({\color{blue}\parbox{0.07in}{$\hspace{-0.0075in}\vspace{-0.01in}\mathsmaller{\square}$}}$\,$); 0.8 ({\color{black}\parbox{0.0825in}{$\hspace{-0.0070in}\vspace{-0.015in}\mathlarger{\mathlarger{\mathlarger{\circ}}}$}}$\,$) times the maximum value computed using the present approach. Results are plotted with the premultiplied spectral density of streamwise velocity fluctuations computed from DNS data. Contours are plotted at 0.2 (\parbox{0.115in}{\color{Red} $\vspace{0.015in} \mathlarger{\mathlarger{\blacktriangle}}$}); 0.5 (\parbox{0.09in}{\color{Blue} $\vspace{0.01in}\mathsmaller{\blacksquare}$}); 
    0.8 (\parbox{0.09in}{ \color{black}$\vspace{-0.01in}\mathlarger{\mathlarger{\mathlarger{\bullet}}}$}) times the maximum value from DNS data at $Re_\tau=934$ \citep{delAlamo2004} (https://torroja.dmt.upm.es/channels/data/spectra/).}}
    \label{fig:energy_spec_dns}
\end{figure}

\begin{figure}
	(a) \hspace{2.6in} (b)	
	
	\centering

	\includegraphics[scale=0.21]{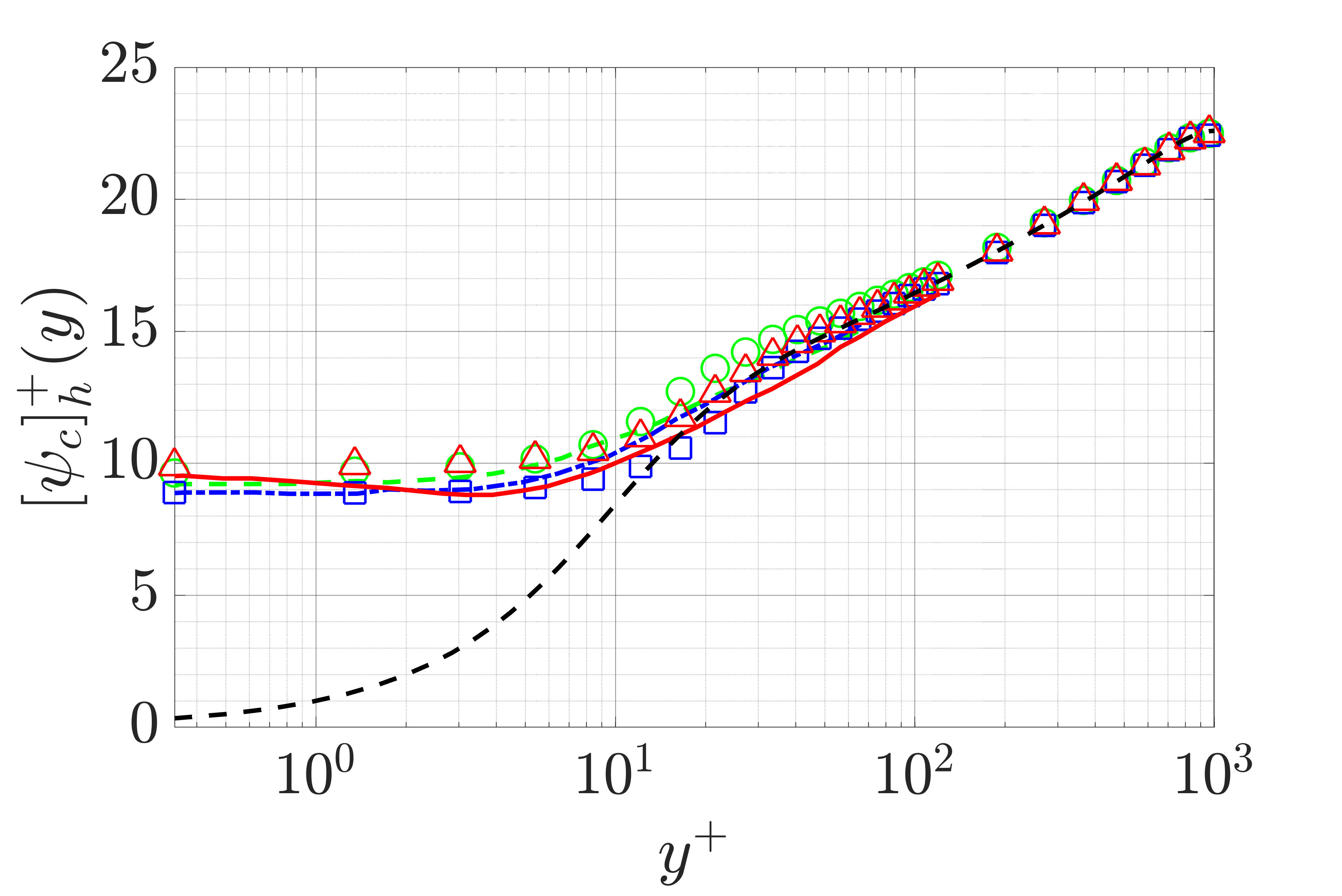}
	\hfill
	\includegraphics[scale=0.17]{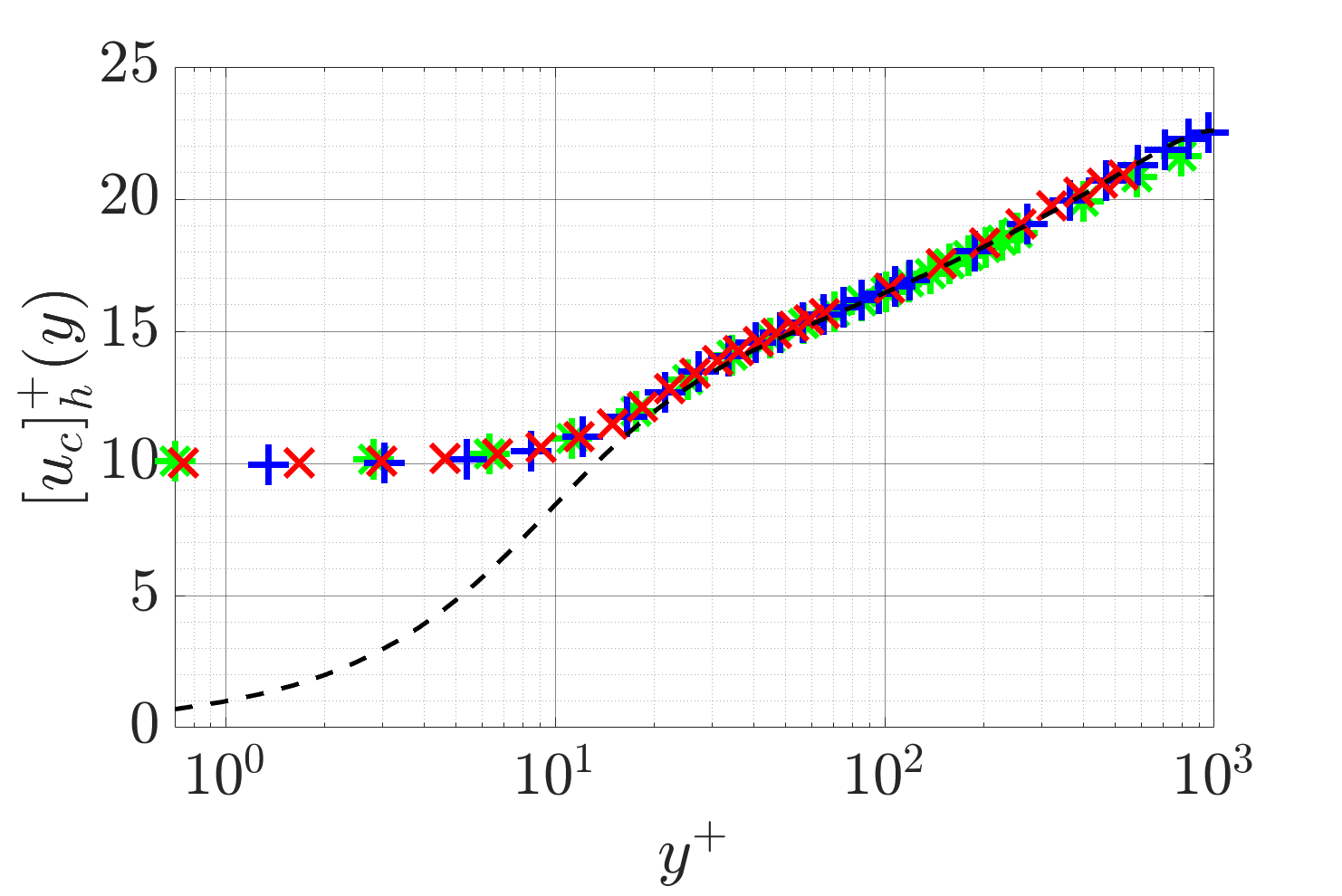}

    \caption{\cliuf{(a) The average convective velocity  of velocity fluctuations, 
	$[\cliu{\psi}_c]_h^+(y)$: $\psi'=u'$ ({\color{red}\parbox{0.0825in}{$\hspace{-0.0070in}\vspace{-0.015in}\mathlarger{\triangle}$}}$\,$); $\psi'=v'$ ({\color{blue}\parbox{0.07in}{$\hspace{-0.0075in}\vspace{-0.01in}\mathsmaller{\square}$}}$\,$); $\psi'=w'$ ({\color{green}\parbox{0.0825in}{$\hspace{-0.0070in}\vspace{-0.015in}\mathlarger{\mathlarger{\mathlarger{\circ}}}$}}$\,$) computed using the \cliuf{present approach} and (\ref{uc1Ddefn}) with their corresponding weighting functions $h=\langle|\mathcal{F}_{xz}(\psi')|^2\rangle k_x^2$ and an averaging domain of $(\lambda_x^+,\lambda_z^+)>(500,80)$ at $Re_\tau=1000$. Results are plotted with convective velocities computed from DNS data \citep{Geng2015} at $Re_\tau=932$: \cliua{$\psi'=u'$} (\parbox{0.115in}{\color{Red}$\mline\mline$ }); \cliua{$\psi'=v'$} (\parbox{0.165in}{\color{Blue} $\dashdot$} );
     \cliua{$\psi'=w'$} (\parbox{0.15in}{ \color{green}$\dashed$}).
	(b) The model-based average convective velocity for streamwise fluctuations \cliua{$\psi'=u'$} computed from (5.1) with the weighting function $h=\langle|\mathcal{F}_{xz}(u')|^2\rangle k_x^2$  over averaging domain $(\lambda_x^+,\lambda_z^+)>(500,80)$ at $Re_\tau=550$ ({\color{red}\parbox{0.0825in}{$\hspace{-0.0070in}\vspace{-0.015in}\mathlarger{\times}$}}$\,$); $Re_\tau=1000$ ({\color{blue}\parbox{0.07in}{  $\hspace{-0.0075in}\vspace{-0.01in}\mathlarger{\mathlarger{+}}$}}$\,$); $Re_\tau=5200$ ({\color{green}\parbox{0.0825in}{$\hspace{-0.0070in}\vspace{-0.015in}${\ding{83}}}}$\,$). The \cliuf{black} dashed lines in both (a) and (b) are the turbulent mean velocity profile at $Re_{\tau}\approx1000$ from \citet{Lee2015}.}}	
	\label{uc1D}
\end{figure}

\cliub{The averaging} domain $\Omega:(\lambda_x^+,\lambda_z^+)>(500,80)$ \cliub{was chosen} \cliua{to include the sublayer streaks proposed as the source of the elevated near-wall convective velocity \citep{Kim1993} but to avoid the} nonlinear effects that dominate at smaller scale\cliub{s}, \cliub{where our linear analysis is not expected to be valid}. \cliuf{The limitations of our input-output based approach at these smaller scales can be understood by examining the energy spectrum that is compared to that of DNS data in figure \ref{fig:energy_spec_dns}. Here it is clear that the spectrum for DNS falls off much faster with decreasing wavelength than the spectrum obtained using the input-output based approach in this work. The relatively heavier weighting at the small scales (wavelength) structures imposed through the present approach results in a lower overall convective velocity \cliua{in these regions.} Our choice of averaging domain eliminates the effect of this heavier weighting \cliub{and implicitly assumes that smaller wavelengths} are energetically negligible. \cliub{Therefore, these small wavelength components do not contribute} to \rev{the average convective velocity computed using our input-output approach.}
}

\cliuf{The performance of the input-output based model at small scales may be improved by integrating known correlations from DNS or experimental data, e.g., shaping the forcing based on  spatial or temporal correlations obtained via simulation data \citep{Moarref2014,Zare2017}. Improvements could also potentially be realized by using an eddy viscosity based enhancement of the LNS equations \citep{reynolds1972mechanics}, which \citet{Zare2017} have shown can provide similar improvements to the input-output response as the introduction of colored-in-time forcing. Understanding the relative benefits of each of these approaches over the current model is a topic of ongoing work.}

Figure \ref{uc1D}(a) compares the resulting \cliub{mean} convective velocities to those obtained from \cliub{the} DNS data \cliub{based computations} of \citet{Geng2015}. \cliub{The plot demonstrates that} the model-based average convective velocities of \cliua{the} streamwise, wall-normal\cliua{,} and spanwise velocity fluctuations \cliub{show good agreement with those} computed from DNS data \citep{Geng2015}.

\cliu{Figure \ref{uc1D}(b) replots the results for the streamwise velocity fluctuations in inner units for} $\Rey_{\tau}=550$, $\Rey_{\tau}=1000$, and $\Rey_{\tau}=5200$. \cliu{The results collapse with} the average convective velocities \cliuf{computed} from the \cliuf{input-output based approach} at different Reynolds numbers all \cliu{tending} to a constant value $\approx10u_{\tau}$ near the wall. \cliu{This} Reynolds number invariance of convective velocities is consistent with the results reported \cliua{in} figure 3 of \citet{Geng2015}.

\cliua{The Reynolds number dependence can be analyzed by rewriting} equations (\ref{NSPFT2}) and (\ref{NSPFT1}) using \cliue{the following change of variables} $k_x=k_x^+Re_{\tau}$, $\hat{\nabla}=\hat{\nabla}^+Re_{\tau}$, $\frac{d\bar{u}}{dy}=Re_{\tau}\frac{d\bar{u}}{dy^+}$ and $\hat{\Delta}=\hat{\Delta}^+Re^2_{\tau}$, which gives:
\cliua{\begin{subequations}
\begin{align}   
\im k_x^+ (\bar{u} - c)\bld{\hat{u}}'
-\hat{\Delta}^+ \bld{\hat{u}}' + \hat{v}'  \dd{\bar{u} }{y^+} \bld{i}
+ \hat{\nabla}^+ \hat{p}' &= \frac{\bld{\hat{f}}'}{Re_{\tau}}   \label{NSPFT2Inner}, \\
\hat{\nabla}^+ \bcdot \bld{\hat{u}}'&=0. \label{NSPFT1Inner} 
\end{align}
\end{subequations}}\cliuf{As} \cliua{neither} the mean velocity profile $\bar{u}$ \cliua{nor} the mean shear $d\bar{u}/dy^+$ at a specific $y^+$ vary over \cliuf{the} Reynolds number in the near-wall region \cliuf{(see e.g., Chapter 7.1.4 of \citet{pope2000turbulent})}, the left hand side of equation (\ref{NSPFT2Inner}) \cliua{does} not \cliua{significantly} vary over Reynolds number. The right hand side of equation (\ref{NSPFT2Inner}) is related to $Re_{\tau}$, but \cliua{the} Reynolds number only influences the magnitude of stochastic forcing $\bld{\hat{f}}'$. According to equation (\ref{phipsi1}) and (\ref{ucdefn}), \rev{the phase speed, $c$, at which $\Phi_{\hat{u}'}$ peaks does not change\cliua{,} and thus\cliua{,} the convective velocity of \cliue{the} streamwise fluctuation\cliua{s}, $\cliu{u}_c$ remains unaffected. This leads to the Reynolds number independence observed in the right panel of figure \ref{uc1D}.

}

This inner units scaling was also previously observed by \citet{Moarref2013} for streamwise energy density and further generalized by \citet{sharma2017scaling} in the framework of resolvent analysis. They end up with \cliua{the same $Re_\tau^{-1}$ scaling for the spatio-temporal transfer function as shown in equation (A4) of \citet{Moarref2013}, and they also pointed out that $Re_\tau$ independence of turbulent mean profile $\bar{u}(y)$ for $y^+\lesssim
 100$ is necessary for this inner units scaling.}

Having validated \cliuf{the ability of the input-output based computations} to reproduce the mean trends, we next investigate the scale dependence of the convective velocity.

\subsection{Scale-by-scale analysis of convective velocity}
\label{ssec:ucScale}

\begin{figure}
\centering 
\begin{subfigure}[b]{\textwidth}
	
 \large{ (a) }
 
\centering

\includegraphics[scale=0.31]{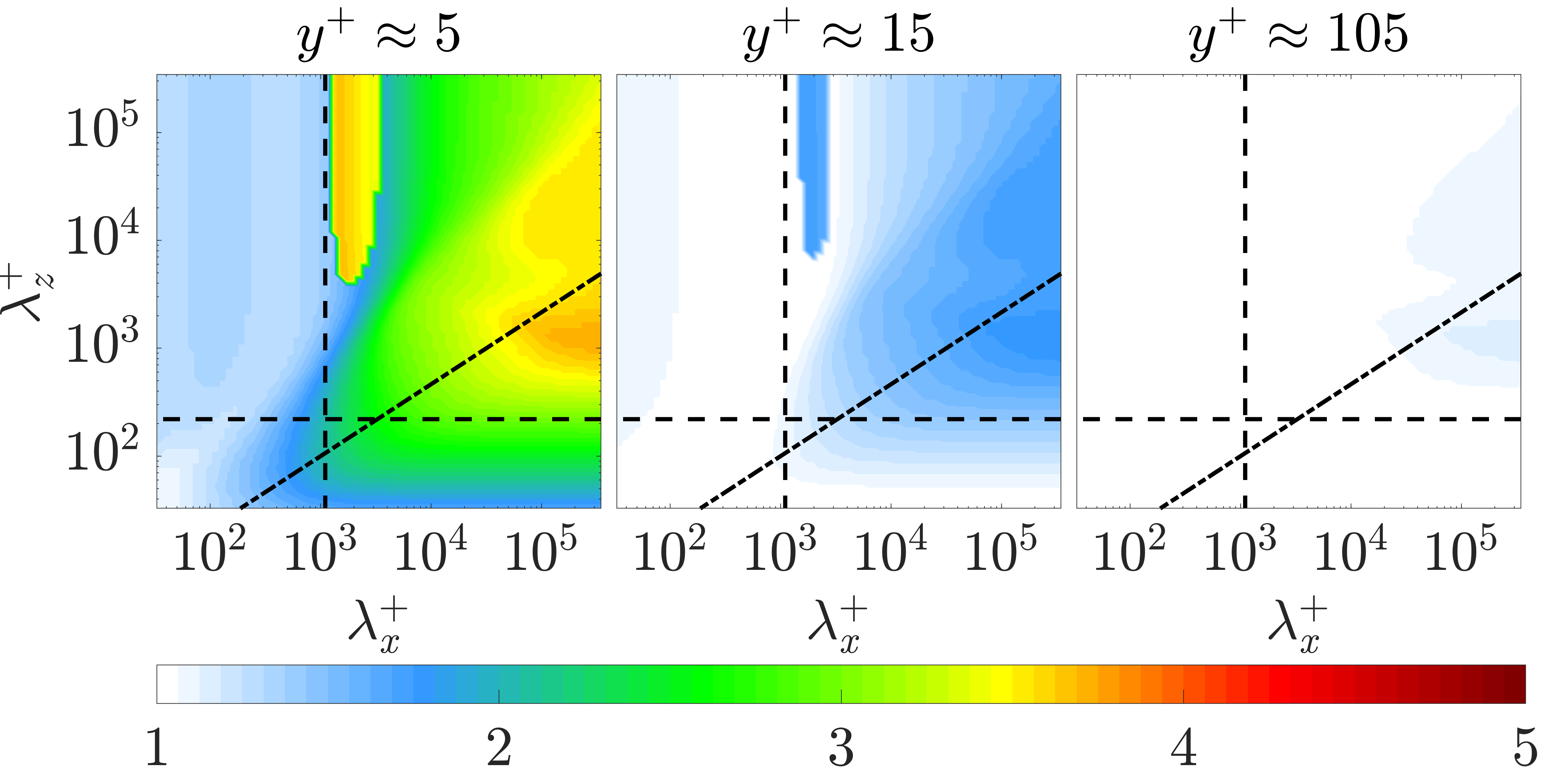}

\end{subfigure}

\begin{subfigure}[b]{\textwidth}
	
\large{ (b) }

\centering

\includegraphics[scale=0.31]{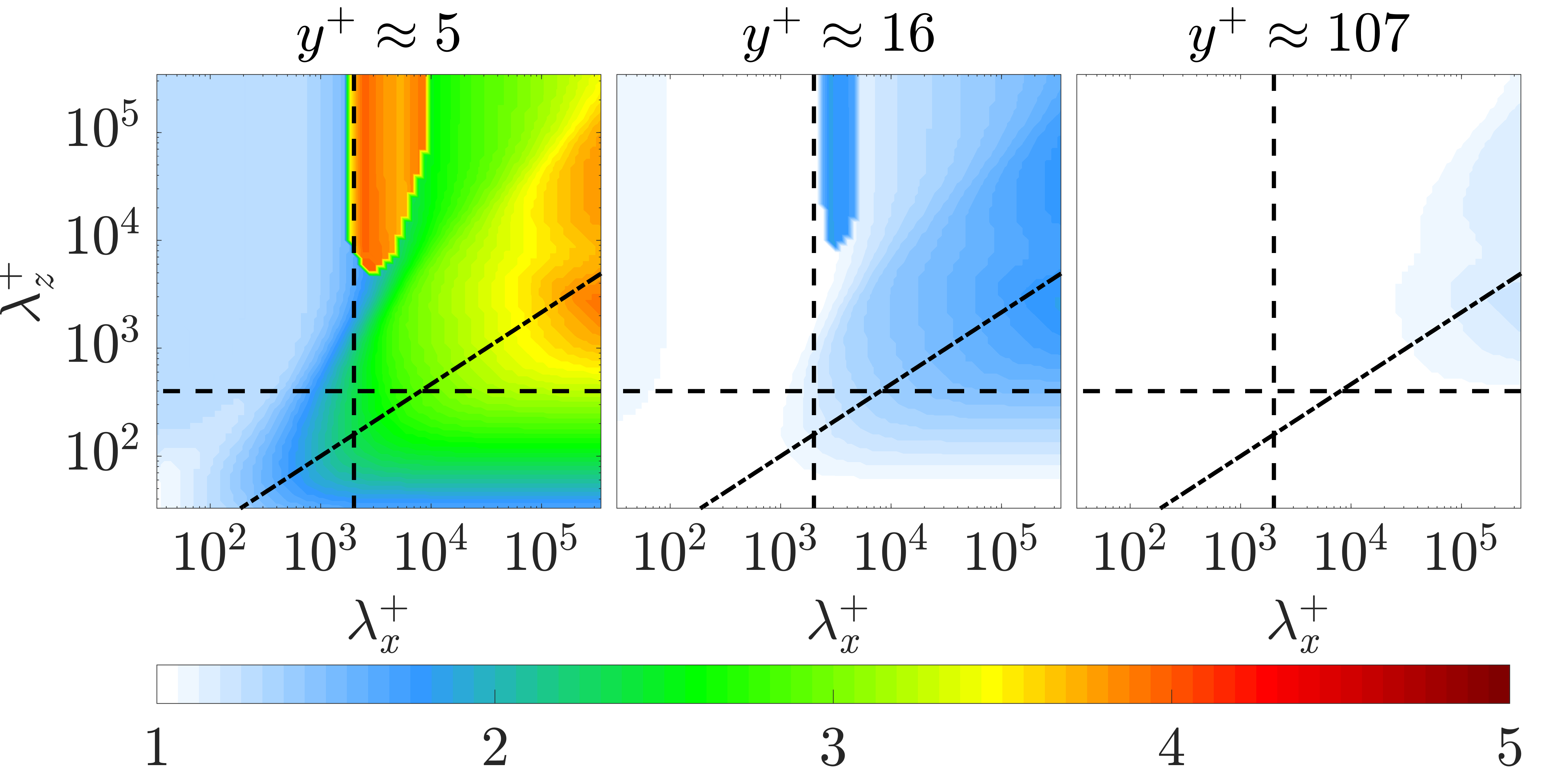}
\end{subfigure}

\begin{subfigure}[b]{\textwidth} 
		\large{ (c) }
		
		\centering
		
		\includegraphics[scale=0.31]{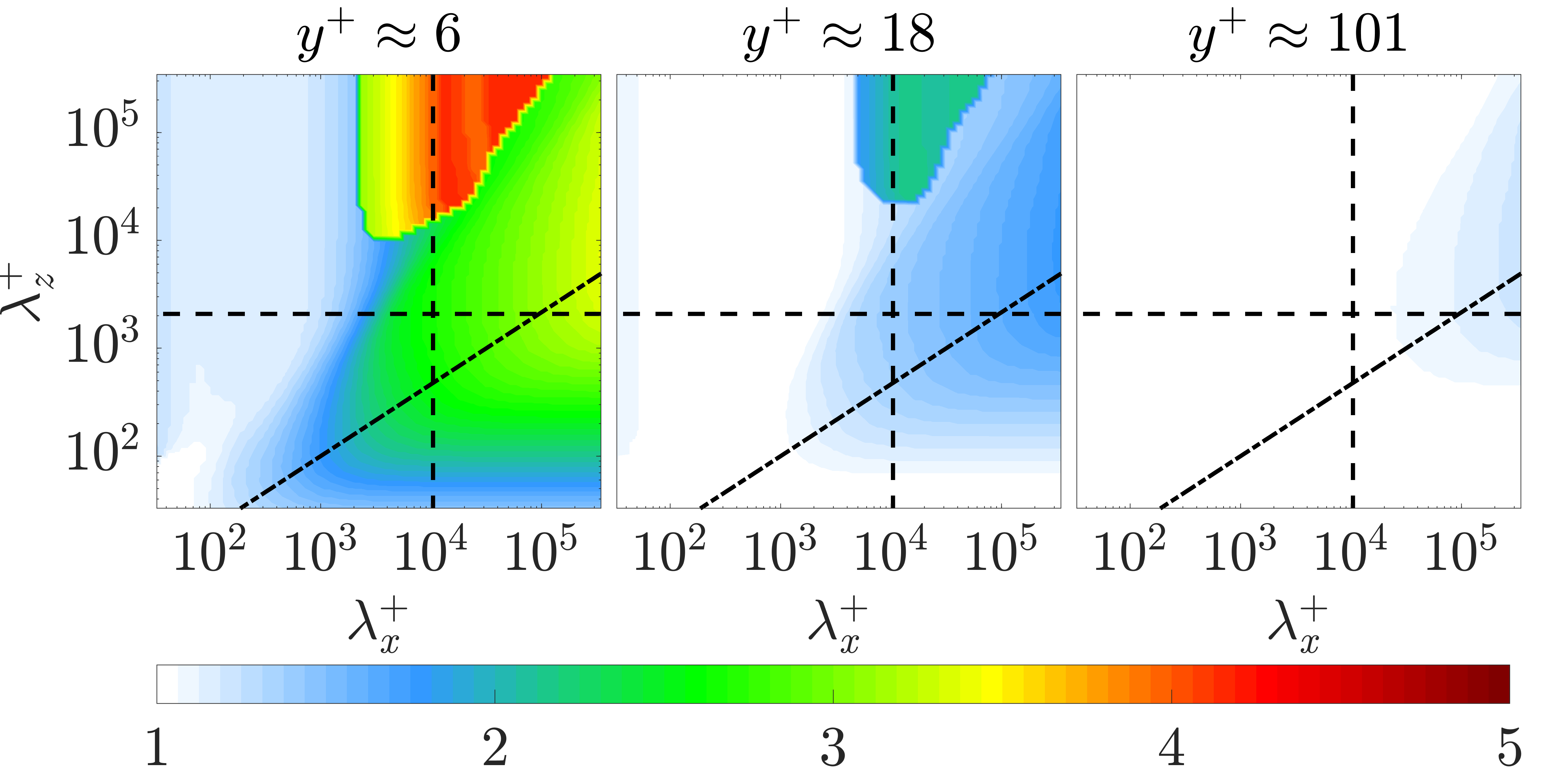}

\end{subfigure} 
\caption{\cliuf{S}cale-dependent convective velocity normalized by the local mean velocity  $\cliu{u}_c(y;\lambda_x,\lambda_z)/\bar{u}(y)$ at (a) $\Rey_{\tau}=550$, (b) $\Rey_{\tau}=1000$\rev{, and
(c) $\Rey_{\tau}=5200$.} 
The black dashed lines are given by $(\lambda_x,\lambda_z) = (2,0.4)$,  which are identified by \citet{delAlamo2009} as the \cliue{lower-bound} of the large-scale convective velocity. The black dash-dot lines are $\lambda_z^+ = {\lambda_x^+}^{\frac{2}{3}}$, which fit through the knee of these contours. }  
	\label{uc_nearWall_tripanel_localmean}
\end{figure}

The model-based approach \cliuf{employed} herein allows one to calculate the contribution of each individual $(\lambda_x,\lambda_z)$ wavelength pair to the local convective velocity at each wall-normal location; i.e., $\cliu{u}_c(y;k_x,k_z)$. We now take advantage of this feature to investigate the scale-dependence of the convective velocity and examine which scales contribute to its deviation from the local mean velocity in various regions of the flow. Figure \ref{uc_nearWall_tripanel_localmean} shows the convective velocity of \cliu{the} streamwise velocity fluctuations normalized by the local mean velocity: $\cliu{u}_c(y;k_x,k_z)/\bar{u}(y)$ for (a) $\Rey_{\tau}=550$, (b) $\Rey_{\tau}=1000$, and (c) $\Rey_{\tau}=5200$, as a function of the streamwise-spanwise wavelengths $(\lambda_x^+,\lambda_z^+)$ at wall-normal locations associated with the viscous sublayer ($y^+\approx 5$), the buffer layer ($y^+\approx 15$), and the log-law region ($y^+\approx 100$). As expected, the results in the viscous sublayer show the most significant deviations from the mean velocity, whereas there is little difference in the log-law region, which is consistent with \cliua{the data} in figure \ref{uc1D}.

In figure \ref{uc_nearWall_tripanel_localmean}, the convective velocities of structures in the intermediate scale range $\lambda_x\approx2$ show \cliua{a discontinuity as the} streamwise wavelength $\lambda_x$ varies. This phenomenon was also observed for the scale-dependent convective velocity of wall-pressure in pipe flow predicted using resolvent analysis with broadband forcing; \cliub{see} figure 12(a) in \citet{Luhar2014}. In the term-by-term analysis in section \ref{sec:term}, we will further confirm that the convective velocity \cliua{of structures associated with these scales} is highly influenced by the pressure.

\begin{figure}
\begin{subfigure}[b]{\textwidth}

\centering

\includegraphics[scale=0.31]{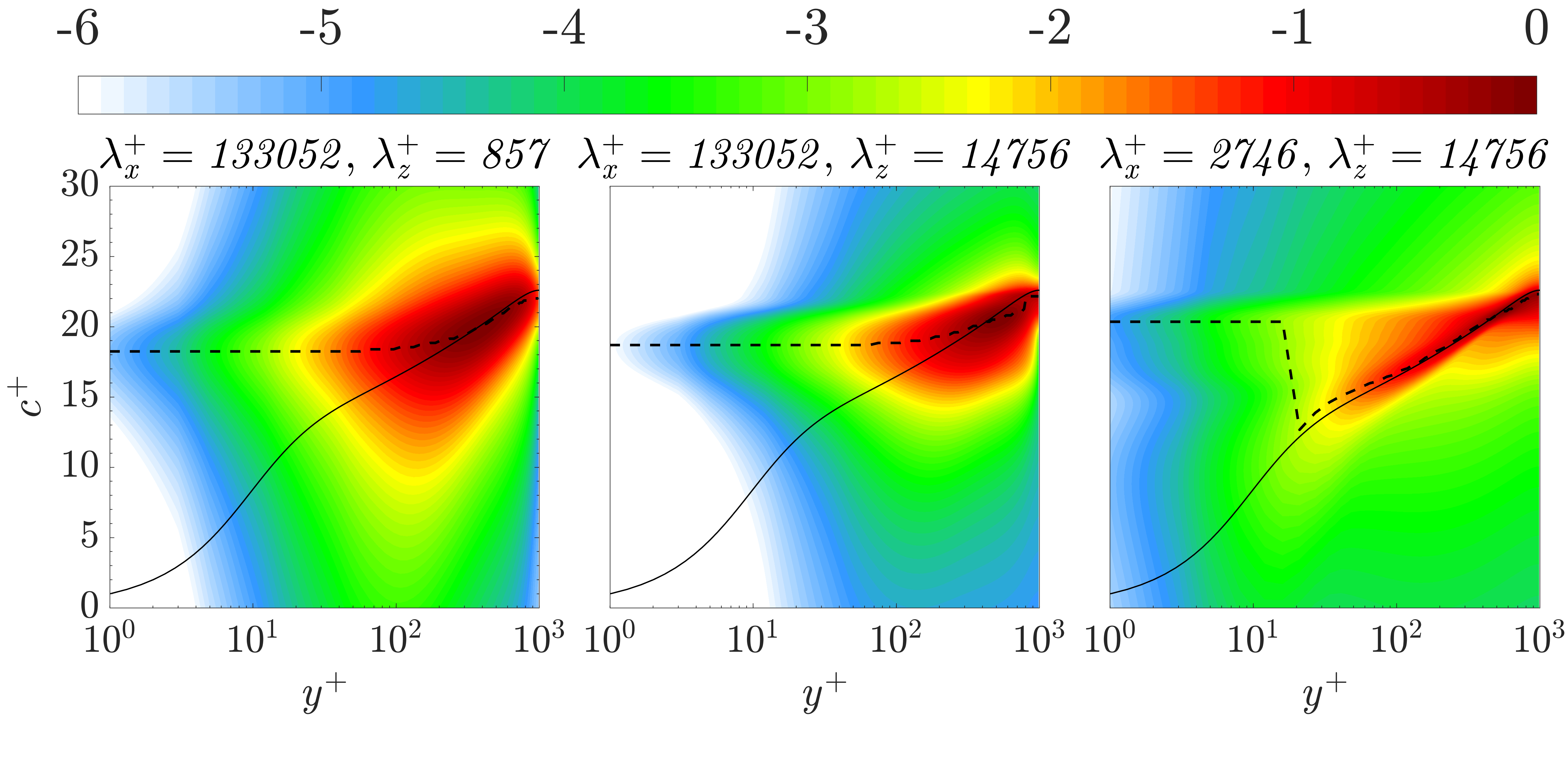}
\end{subfigure}

\hspace{2.4cm} \large{ (a) }\hspace{3.3cm}
\large{ (b) }\hspace{3.3cm}
\large{ (c) }

\begin{subfigure}[b]{\textwidth} 
		\includegraphics[scale=0.68]{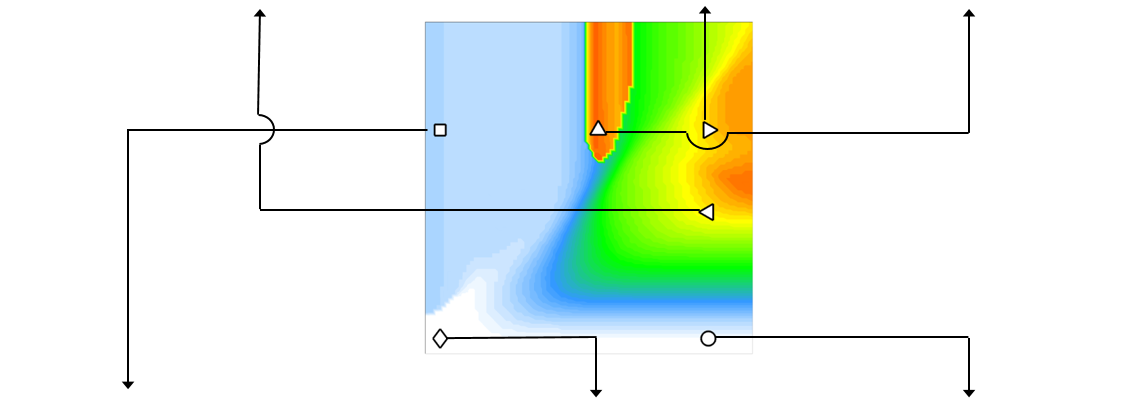}
		
\end{subfigure} 
	
\hspace{2.4cm} \large{ (d) }\hspace{3.3cm}
\large{ (e) }\hspace{3.3cm}
\large{ (f) }

\begin{subfigure}[b]{\textwidth} 
		\centering
		
		\includegraphics[scale=0.31]{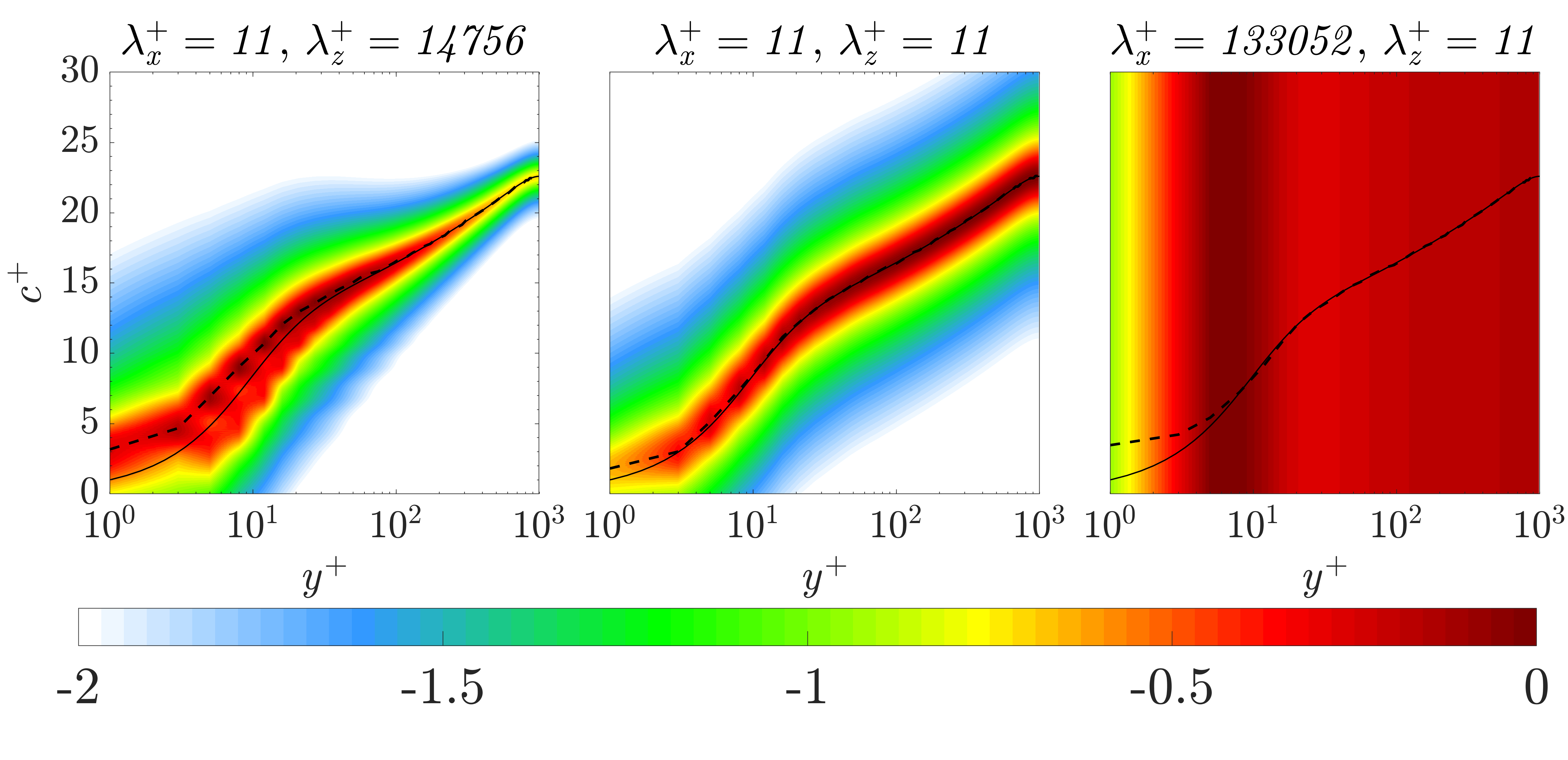}
		
\end{subfigure} 
	\caption{\cliuf{Normalized} power spectral density of streamwise velocity fluctuations $\frac{\Phi_{\hat{u}'}(y;k_x,k_z,c)}{\max\limits_{c,y}(\Phi_{\hat{u}'}(y;k_x,k_z,c))}$ over wall-normal location $y^+$ and phase speed \cliuf{$c^+$}. \cliue{The symbols represent locations associated with large-scale structures at} (a) $\vartriangleleft (\lambda_x^+,\lambda_z^+)=(133052,857)$ and (b) $\vartriangleright (\lambda_x^+,\lambda_z^+)=(133052,14756)$; (c) intermediate-scale structures  $\triangle \; (\lambda_x^+,\lambda_z^+)=(2746,14756)$, and structures with small streamwise or spanwise wavelengths, respectively: (d) $\square\; (\lambda_x^+,\lambda_z^+)=(11,14756)$, (e) $\mathlarger{\mathlarger{\mathlarger{\diamond}}}\; (\lambda_x^+,\lambda_z^+)=(11,11)$, and (f) $\mathlarger{\mathlarger{\mathlarger{\circ}}}\; (\lambda_x^+,\lambda_z^+)=(133052,11)$. The color is in base 10 logarithmic scale. The black solid lines represent the mean streamwise velocity profile and the black dashed lines are convective velocites computed using the method in section \ref{sec:method}. The middle panel, which is \cliue{reproduced from} figure \ref{uc_nearWall_tripanel_localmean}(b) at $y^+\approx5$ for $Re_\tau=1000$,  \cliue{indicates the locations corresponding to each symbol.}
	}
	
	\label{energy}
\end{figure}

Figure \ref{energy} shows the PSD, $\Phi_{\hat{u}'}(y;k_x,k_z,c)$ computed from equation (\ref{phipsi1}), \cliua{of the streamwise velocity fluctuations for} $Re_\tau=1000$ \cliua{as a function of} wall-normal location $y^+$ and phase speed $c^+$ at six different streamwise--spanwise wavelength pairs. Figures \ref{energy}(a) and (b) show that \cliua{the energy of the} large wavelength structures at example points $\vartriangleleft (\lambda_x^+,\lambda_z^+)=(133052,857)$ and $\vartriangleright (\lambda_x^+,\lambda_z^+)=(133052,14756)$, are concentrated near $(y^+, c^+)\approx(200,18.4)$, \cliua{and that} structures traveling at $c^+\approx 18.4$ \cliuf{provide the largest contribution to the spectral density in the near-wall region}. \cliuf{Figure 5(c) indicates} that structures traveling at a higher velocity than the local mean also contribute most to the PSD in the near-wall region \cliuf{for the intermediate-scale structures}. In contrast, \cliue{the} PSD distributions over ($c^+,y^+$) \cliue{for} \cliua{structures with small \cliue{streamwise} wavelengths} are more concentrated near the mean velocity profile $\bar{u}$ as shown in figures \ref{energy}(d), and (e) for example points $\square\; (\lambda_x^+,\lambda_z^+)=(11,14756)$, and $\mathlarger{\mathlarger{\mathlarger{\diamond}}}\; (\lambda_x^+,\lambda_z^+)=(11,11)$.

\cliuf{Figure \ref{energy} indicates that the PSD distribution over phase speed is nearly symmetric about its peak in figures \ref{energy}(a), (b), (d), and (e), which indicates close correspondence between the center of gravity and peak of the PSD definitions of convective velocity. For the representative intermediate flow scale plotted in figure \ref{energy}(c), the PSD distribution over the phase speed shows skewness, which is expected to lead to differences in the convective velocity obtained by considering the distribution. In order to evaluate the differences, we recomputed the results in figure \ref{uc_nearWall_tripanel_localmean}(b) using a center of gravity definition $\check{\psi}_c(y;k_x,k_z)\equiv\frac{\int c\Phi_{\hat{\psi}'}(y; k_x, k_z, c)dc}{\int\Phi_{\hat{\psi}'}(y; k_x, k_z, c)dc}$ and found that there are indeed differences for these scales. Specifically, the discontinuity in the convective velocity as streamwise wavelength $\lambda_x$ varies over larger $\lambda_z$ is smoothed. Differences also occur at the flow scales that are very small in the spanwise direction but very long in the streamwise direction; e.g., the structures indicated by $\mathlarger{\mathlarger{\mathlarger{\circ}}}\; (\lambda_x^+,\lambda_z^+)=(133052,11)$ in figure \ref{energy}(f). These differences are not surprising because figure \ref{energy}(f) demonstrates that the PSD is quite flat and therefore advection does not dominate at these flow scales. Here, neither definition of convective velocity is physically meaningful, as the maximum is not associated with a clear peak and the center of gravity definition merely selects the center point of the computational domain. The comparison verified that the overall trends that are highlighted in this manuscript, such as the influence of the large-scale structures in the near-wall and buffer regions as well as the slope of the knee through the contours indicated by dash-dot lines in the panels of figure 4 were unchanged when we used the center of gravity in our computations. We therefore proceed with the definition in terms of the peak of the PSD in equation (\ref{ucdefn}) in the remainder of the manuscript.}

\cliub{Both figures \ref{uc_nearWall_tripanel_localmean} and \ref{energy} indicate that large channel spanning structures have an influence on the convective velocity in both the viscous sublayer and the buffer layer.} \cliua{This phenomenon was investigated by \citet{delAlamo2009}, who identified large scales as structures of size $(\lambda_x,\lambda_z)>(2,0.4)$.} \cliub{This large-scale cut-off is identified by horizontal and vertical dashed lines in each panel of figure \ref{uc_nearWall_tripanel_localmean}.}

\begin{figure}
	(a) \hspace{2.6in} (b)	
	
	\centering
    \includegraphics[scale=0.21]{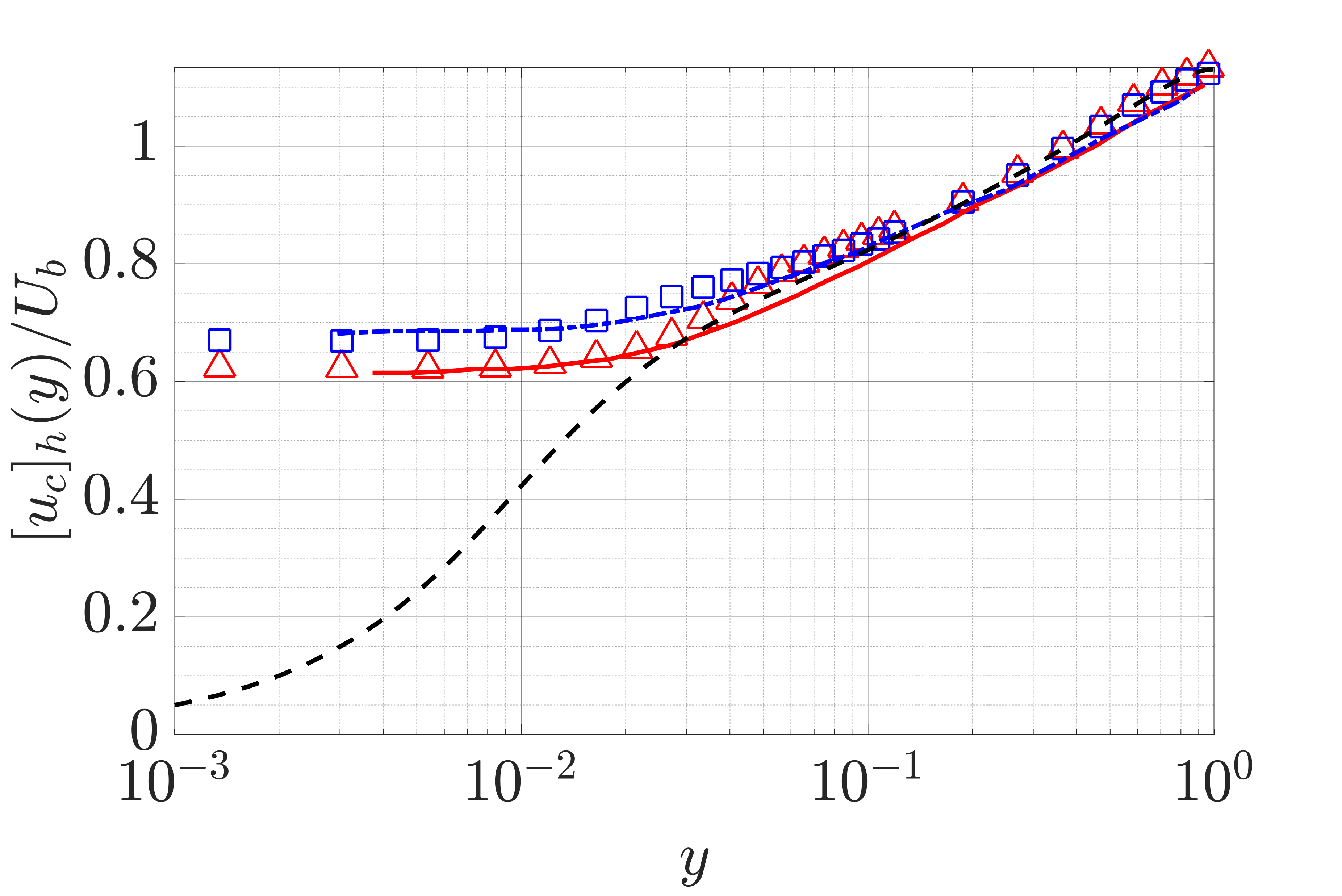}
    \includegraphics[scale=0.25]{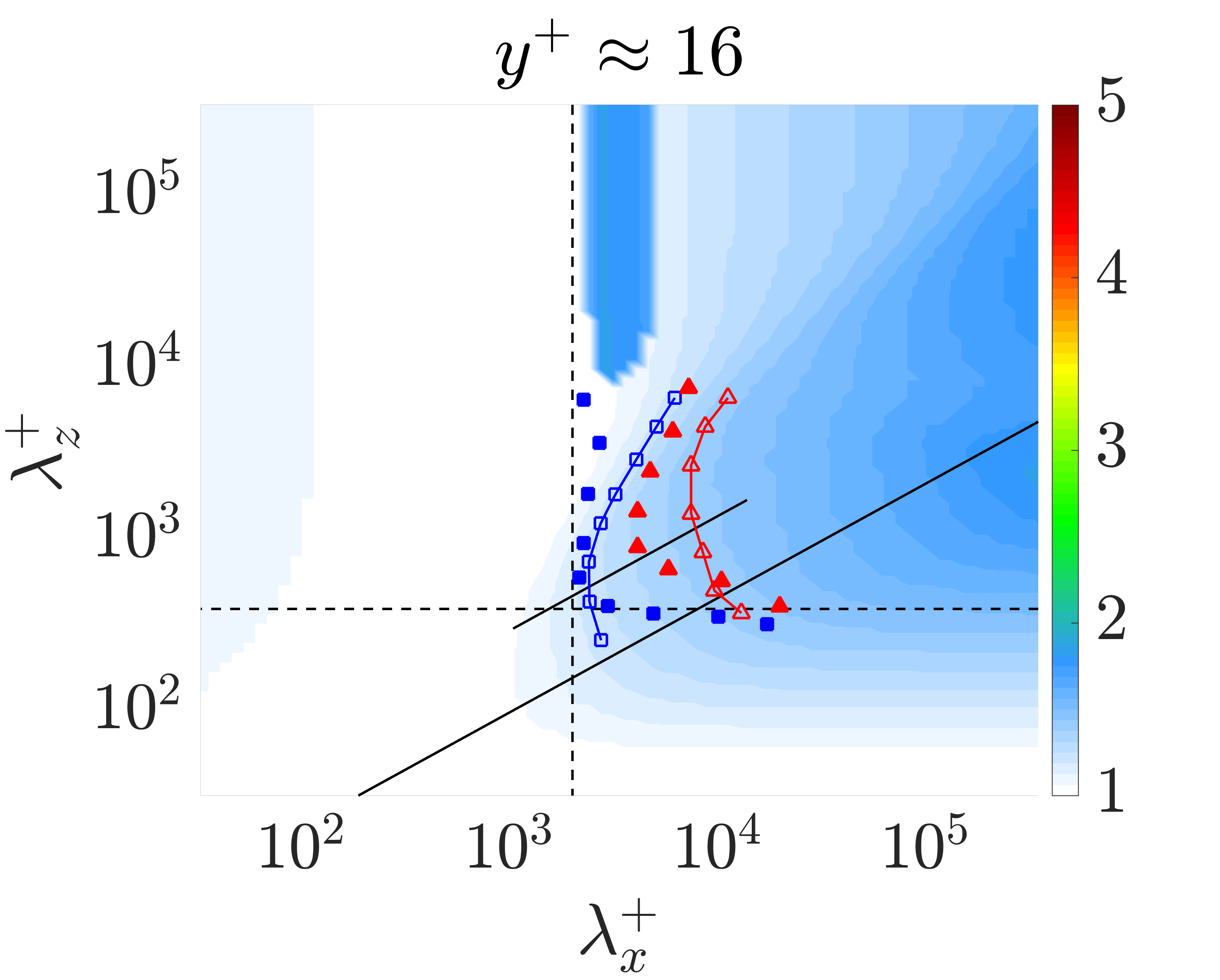}

	\caption{(a) The average convective velocity of streamwise fluctuations $[\cliu{u}_c]_h/U_b$ over $(\lambda_x,\lambda_z)>(2,0.4)$. Model based results at $Re_{\tau}=550$ ({\color{red}\parbox{0.1025in}{$\hspace{-0.0070in}\vspace{-0.015in}\mathlarger{\triangle}$}}) and
			$Re_{\tau}=1000$ ({\color{blue}\parbox{0.07in}{ $\hspace{-0.0075in}\vspace{-0.01in}\mathsmaller{\square}$}}) are compared to convective velocities computed from DNS data \citep{delAlamo2009} at $Re_{\tau}=547$ (\parbox{0.115in}{\color{Red} $\mline\mline$}) and  
	$Re_{\tau}=934$ (\parbox{0.17in}{\color{Blue} $\dashdot$} ). \cliu{For direct comparison with \citet{delAlamo2009}, the results in (a) are scaled by the bulk velocity; i.e., $U_b=\frac{1}{2}\int_{-1}^1\bar{u}(y)dy$.} The \cliuf{black} dashed line is the mean velocity profile at $Re_{\tau}\approx1000$ from \citet{Lee2015}. \cliu{(b) Model based scale-dependent convective velocity \cliua{at $y^+\approx 16$ for $Re_\tau=1000$}: contour lines $u_c(y;k_x,k_z)/\bar{u}(y)=1.40$ ({\color{red}$\mline\mline$\parbox{0.0825in}{$\hspace{-0.0070in}\vspace{-0.015in}\mathlarger{\triangle}$}$\mline\mline$});
		$u_c(y;k_x,k_z)/\bar{u}(y)=1.21$ ({\color{blue}$\mline\mline$\parbox{0.07in}{ $\hspace{-0.0075in}\vspace{-0.01in}\mathsmaller{\square}$}$\mline\mline$}); are compared to convective velocities computed from DNS data \cliua{at $y^+=15$ and $Re_\tau=934$} \citep{delAlamo2009} $u_c(y;k_x,k_z)/\bar{u}(y)=1.40$ (\parbox{0.115in}{\color{Red} $\vspace{0.015in} \mathlarger{\mathlarger{\blacktriangle}}$});  
	$u_c(y;k_x,k_z)/\bar{u}(y)=1.21$ (\parbox{0.09in}{\color{Blue} $\vspace{0.01in}\mathsmaller{\blacksquare}$}). The black dashed lines are given by $(\lambda_x,\lambda_z) = (2,0.4)$. The black solid lines are $\lambda_z^+ \sim {\lambda_x^+}^{\frac{2}{3}}$, which fit through the knee of model based convective velocity and DNS data \cliua{contours} \citep{delAlamo2009}, respectively. }}
	\label{uc1DReynoldsMean}
\end{figure}

\cliub{Figure \ref{uc1DReynoldsMean}(a) plots the mean convective velocities obtained for the averaging domain $(\lambda_x,\lambda_z)>(2,0.4)$} for $Re_{\tau}=550$ and $Re_{\tau}=1000$. \cliub{A comparison to DNS data indicates that the model produces} qualitatively and quantitatively similar behavior to results computed from DNS data (figure 5(a) in \citet{delAlamo2009}). \cliub{Note the results in figure \ref{uc1DReynoldsMean}(a)} are scaled \cliub{by} the bulk velocity; i.e., $U_b=\frac{1}{2}\int_{-1}^1\bar{u}(y)dy$ for direct comparison with \citet{delAlamo2009}.

\cliub{In figure \ref{uc1DReynoldsMean}(b), we further analyze }\cliu{\cliua{the influence of the large scales by comparing} the scale-dependent convective velocity \cliua{at $y^+\approx 16$ with $Re_\tau=1000$} from our \cliuf{approach} to figure 3(a) in \citet{delAlamo2009}. \cliub{\cliuf{The darker blue region of the largest $\lambda_x^+$ and moderate to largest $\lambda_z^+$ in figure \ref{uc1DReynoldsMean}(b) indicates the influence of large and very large-scale motion on the convective velocity in this region,} which supports previous studies \citep{Kim1993,Krogstad1998} indicating that fast moving structures centered further away from the wall (where the local mean velocity is larger) have an influence very near the wall due to their large size \citep{Dinkelacker1977,Kreplin1979,Farabee1991,Kim1993,Hutchins2011}.}}

\cliua{A linear fit through the knee of the contour plot in figure \ref{uc1DReynoldsMean}(b) shows self-similar structures with a ratio $\lambda_z^+ \sim {\lambda_x^+}^{\frac{2}{3}}$ \cliub{in both our results and} \cliue{those} reported in \citet{delAlamo2009}.} \cliuf{
This type of $x$--$z$ similarity in energy spectra density has been previously observed in the context of geometric self-similarity \citep{delAlamo2004,chandran2017two}.} For example, del \'{A}lamo and co-authors \citep{delAlamo2003,delAlamo2004} found that the isocontours of the pre-multiplied \rev{energy} spectrum of $u'$ form a corner \cliua{centered} along $\lambda_z  \sim \lambda_x^{n}$ with  $\frac{1}{3} < n < 1$. \cliua{In that work, the} value of $n$ changed over \cliua{the} wall-normal location with \cliue{lower bound on $n$} near the buffer layer ($y^+\approx 15$), increasing to $n=\frac{1}{2}$ in the log-law region and reaching $n=1$ in the outer region of the flow. Recent experimental measurements of two-dimensional spectra in zero-pressure gradient boundary layers \cliuf{indicate a $\lambda_z \sim \sqrt{\lambda_x}$ ($Re_\tau= 2430$) and $\lambda_z\sim \lambda_x$ ($Re_\tau=26090$)} relationship in the start of the log-law \cliua{region} \citep{Chandran2016,chandran2017two}. 

\cliuf{The scaling law of convective velocity explored here is closely related to the temporal self-similarity previously observed in the literature. More specifically, \citet{lozano2014time} showed that tall attached structures are both geometrically and temporally self-similar with lifetimes proportional to their distance from the wall. They also attribute the lifetime and deformation of these structures to the vertical gradient of their convective velocity. Long lifetimes, which require low dispersion, have been associated with coherent structures \citep{Adrian2007}; e.g., hairpin vortices that are observed to propagate downstream with small velocity dispersion \citep{adrian2000vortex}. Non-dispersive coherent structures are implied by the isocontour lines of the scale dependent convective velocities in figures \ref{uc_nearWall_tripanel_localmean} and \ref{uc1DReynoldsMean}(b), which forms a $\lambda_z^+ \sim {\lambda_x^+}^{\frac{2}{3}}$ knee. Based on this observation, we conjecture that the $x$--$z$ similarity observed here is closely related to the scaling laws of energy spectra.}

To explain the $\lambda_z^+ \sim (\lambda_x^+)^{2/3}$ scaling, \cliua{we employ} a simple model involving the bending of streamlines in the cross-plane due to the presence of a streamwise vortex. This simple model was originally proposed by \citet{Jimenez2004} to explain the contribution of high-momentum streaks to the \rev{energy} spectrum. Consider convection of $\hat{u}'$ due to a point vortex with circulation $\gamma$ in the cross-plane $\p_t u' + \frac{2\pi \gamma}{r^2}\frac{\p u'}{\p \theta} = 0$ where $(r,\theta)$ is the polar representation of the $(y,z)$ plane, \cliua{centered} on the vortex. For a homogeneous shear initial condition, $u'(t=0,r,\theta) \sim Sy = Sr\sin \theta$, with shear rate $S$, we have $u'(t,r,\theta) \sim Sr\sin(\theta - \gamma t/2\pi r^2)$. At a given time, $t$, setting $\theta = \pi$ yields the `size' of the vortex-distorted region as\cliua{:}
\begin{align}
R_z = \sqrt{\gamma t/2\pi^2}. \label{Rzvortex}
\end{align}

Moreover, the length of streak is determined by the velocity difference between its top and bottom with shear rate $S$, which can be roughly estimated as:
\begin{align}
\lambda_x=SR_zt, \label{lowshear}
\end{align}
\cliub{if we assume} that the streak height is roughly equal to its width. We estimate the meandering magnitude of streaks \cliua{by approximating} the spanwise drift of the vortices under the induction of their reflected images across the wall\cliua{, which leads to}
\begin{align}
\lambda_z=2\sqrt{2} w't, \label{meander}
\end{align}
where $w'$ \cliua{denotes a spanwise velocity} fluctuation of the order of $w'^+\approx1$ \citep{kim1987turbulence}. Combining equations (\ref{Rzvortex}), (\ref{lowshear}), and (\ref{meander}) gives the scaling:
\begin{align}
\lambda_z^+\sim(\lambda_x^+)^\frac{2}{3}. \label{23scaling}
\end{align}

Although \cliub{this is an idealized analysis, it leads to the trends observed both here and in DNS based convective velocity analysis. The assumptions underlying this scaling are also consistent with the existence of structures at a wide variety of scales extending} into the channel; i.e., structures reminiscent of Townsend's attached eddies \citep{Townsend1976,perry1986theoretical,marusic2019attached}. \cliua{We} next calculate the wall-normal coherence of these structures to \cliue{further examine this connection.}

\section{Wall-normal coherence of viscous sublayer structures}
\label{sec:coher}
\rev{The convective velocity at each $(\lambda_x,\lambda_z)$ wavelength pair,} $\cliu{u}_c(y;k_x,k_z)$\rev{,} is obtained via the maximization in (\ref{ucdefn}), \cliue{so} we refer to the spectral component of $u'$ defined by $(\lambda_x,\lambda_z,u_c)$ as the `characteristic structure'.  \rev{\cliub{We hypothesize that the} characteristic structure at a given wall-normal location, $y_0$, is responsible for the dominant convection at that location, \rev{and that it also contributes to the energetics elsewhere in the channel \cliua{due to its} wall-normal extent}. In this section, we investigate \rev{this} wall-normal extent \rev{using the spectral coherence between signals at two different wall-normal locations. We focus on the characteristic structures that provide the dominant convection in the viscous sublayer and on wavelength pairs along the knee \cliua{of} the isocontours of $\cliu{u}_c$ in figure \ref{uc_nearWall_tripanel_localmean}; i.e., along $\lambda_z^+ = (\lambda_x^+)^{\frac{2}{3}}$.}}

\begin{figure}
\centering
\includegraphics[scale=0.29]{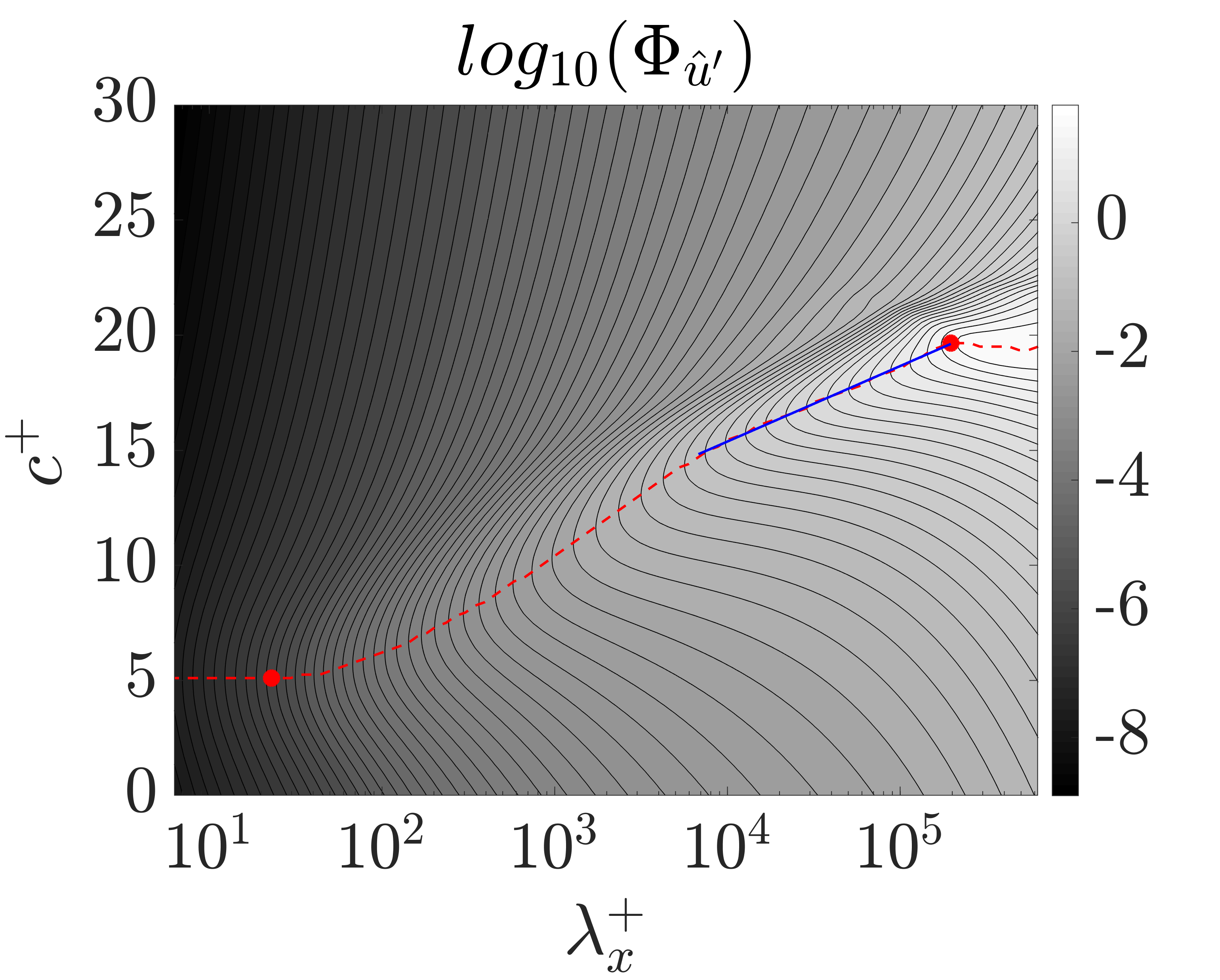}
\caption{Isocontours of $\Phi_{\hat{u}'}$ in the viscous sublayer ($y^+ \approx 5$) along $\lambda_z^+ = (\lambda_x^+)^{\frac{2}{3}}$ at $Re_{\tau}=1000$, calculated using (\ref{phipsi1}). The red dashed line indicates the $c^+$ at which $\Phi_{\hat{u}'}$ peaks for each $\lambda_x^+$, and the blue solid line indicates the region of logarithmic increase. It therefore defines the convective velocity $\cliu{u}_c$ as in (\ref{ucdefn}). 
The markers (\parbox{0.09in}{ \color{red}$\vspace{-0.01in}\mathlarger{\mathlarger{\bullet}}$}) indicate the locations where we evaluate the two-point wall-normal coherence in figure \ref{2D_coherence_uvw_sqrtlambdaz}.}
\label{uc_phi_monopanel}
\end{figure}

The $u'$ \rev{frequency--wavenumber} spectrum, $\Phi_{\hat{u}'}=\langle|\hat{u}'|^2\rangle$ of streamwise fluctuations in the viscous sublayer ($y^+ \approx 5$) for $Re_\tau=1000$, is shown in figure \ref{uc_phi_monopanel} \rev{in terms of phase speed and} wavelengths along $\lambda_z^+ = (\lambda_x^+)^{\frac{2}{3}}$. The autocorrelation maxima defining the convective velocity are plotted as a dashed line. \cliua{The ridge corresponding to \cliue{these} maximum values} asymptote to constants at both the large and small wavelength limits, but show a region of linear growth followed by a region of logarithmic increase (blue solid line) \cliue{between two red circle markers}. The logarithmic behavior is similar to the variation of the mean velocity profile $\bar{u}(y)$ with $y$ and is consistent with the assumption that the dominant viscous sublayer convection at streamwise wavelength $\lambda_x$ arises due to a structure advecting at the local mean velocity at $y \sim \lambda_x^n$ for some $n>0$.

\begin{figure}

\centering 

\begin{subfigure}[b]{\textwidth}

\centering
\includegraphics[scale=0.29]{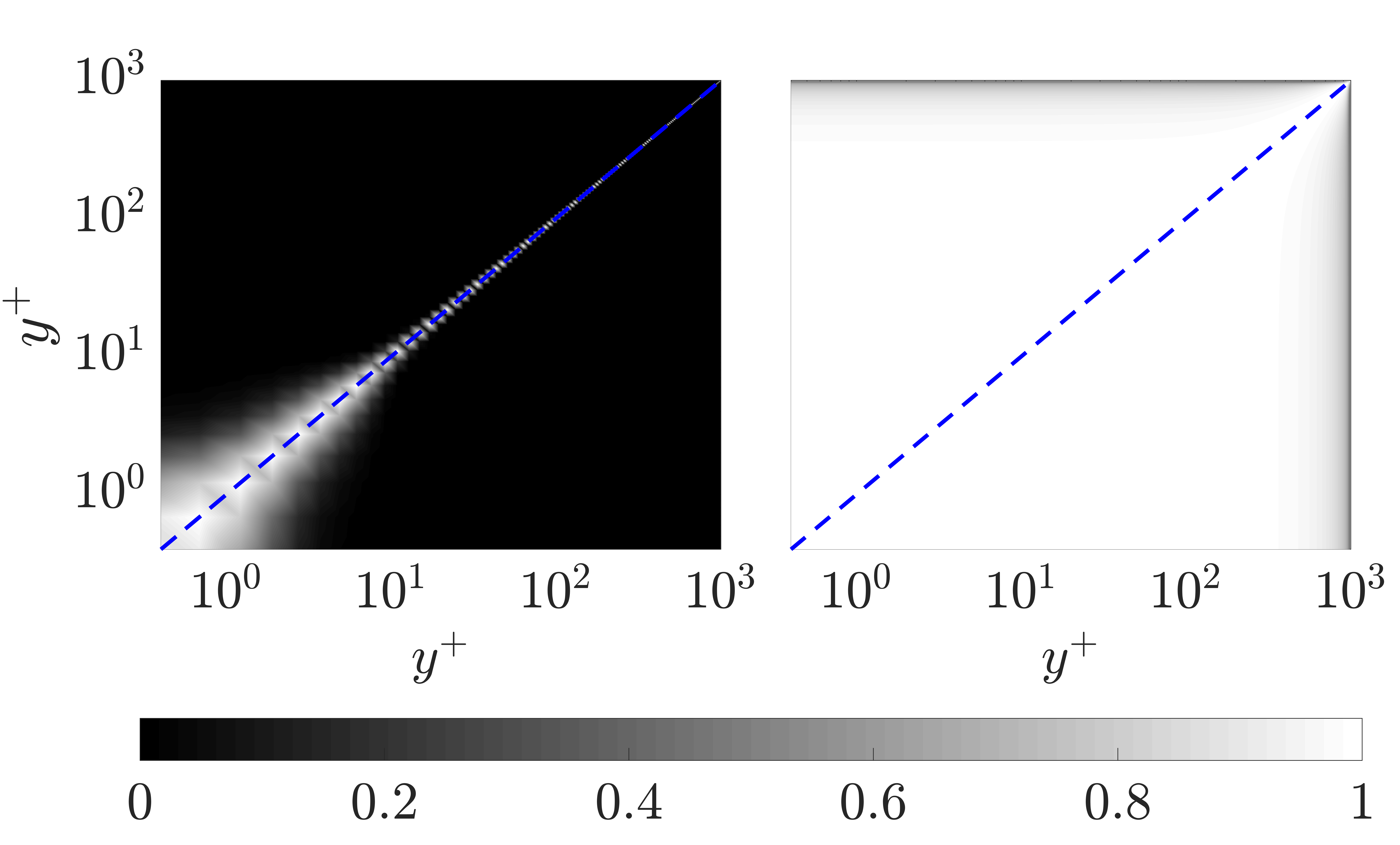}

\end{subfigure}

\caption{Two-point spectral coherence of streamwise velocity fluctuations $\chi_{u'; y'y}$ for data at $Re_\tau=1000$, as defined in (\ref{SpecCohdefn}) at $(\lambda_x^+,\lambda_z^+) \approx (22,8)$ (left) and   $(\lambda_x^+,\lambda_z^+) \approx (2\times 10^5,3\times 10^3)$ (right) indicated by circle markers in figure \ref{uc_phi_monopanel}. Both points are along $\lambda_z^+ = (\lambda_x^+)^{\frac{2}{3}}$, and their phase speeds in friction units are approximately $c^+\approx 5$ for the small-scale structure and $c^+\approx 20$ for the large one. Perfectly coherent signals have a spectral coherence of $1$, and incoherent signals have a spectral coherence of $0$.}
\label{2D_coherence_uvw_sqrtlambdaz}
\end{figure}

The spectral coherence between two wall-normal locations $y'$ and $y$, defined for a \cliua{fluctuating} variable $\psi$ \cliue{is defined }as:
\begin{align}
0 \leq \chi_{\psi'; y'y}(k_x,k_z,c) \equiv \frac{|\Phi_{\hat{\psi}',\text{cross}}(y',y; k_x, k_z, c)|^2}
{\Phi_{\hat{\psi}'}(y'; k_x, k_z, c)\Phi_{\hat{\psi}'}(y; k_x, k_z, c)} \leq 1,
\label{SpecCohdefn}
\end{align} 
where $\Phi_{\hat{\psi}',\text{cross}}$ is the cross-spectral density of $\psi'$ between locations $y'$ and $y$; i.e.,
\begin{align}
\Phi_{\hat{\psi}',\text{cross}}(y',y; k_x, k_z, c) = \langle\hat{\psi}'^*(y'; k_x, k_z, c) \hat{\psi}'(y; k_x, k_z, c)\rangle.
\end{align}
The cross-spectral densities are the off-diagonal components of the matrix obtained from the finite-dimensional representation (\ref{phipsiFD}) of $\Phi_{\hat{\psi}'}$. Figure \ref{2D_coherence_uvw_sqrtlambdaz} shows the two-point spectral coherence for streamwise velocity fluctuations $\chi_{u'; y'y}$ for two characteristic structures along $\lambda_z^+ = (\lambda_x^+)^{\frac{2}{3}}$ in the near-wall region: (a) a short wavelength component,  $(\lambda_x^+,\lambda_z^+) \approx (22,8)$, and (b) a long wavelength component,  $(\lambda_x^+,\lambda_z^+) \approx (2\times 10^5,3\times 10^3)$. The phase speeds associated with these \cliuc{characteristic} structures are indicated in figure \ref{uc_phi_monopanel} by circle markers. The shorter wavelength component is associated with the smaller convective velocity and the longer wavelength component with the larger convective velocity.

In figure \ref{2D_coherence_uvw_sqrtlambdaz}, we see a wall-normal coherence that extends \emph{from} the wall; i.e., it does not involve any wall-detached patches, consistent with Townsend's attached-eddy hypothesis \citep{Townsend1976,perry1986theoretical,marusic2019attached}. As predicted by \cite{delAlamo2009}, the long wavelength component is more coherent further into the channel towards the core than the short wavelength component with its coherence falling to $0$ in the core. This growth of coherence away from the wall with increasing wavelength suggests that the structures contributing to the convective velocity in the viscous sublayer extend from the wall deep into the log-law region, but only weakly into the wake region, reminiscent of the long meandering structures in the log-law region \cliua{whose footprint extends to the} near-wall region \citep{Jimenez2004,Hutchins2007,monty2007large,Guala2006,Balakumar2007}.

Calculations (not presented here for brevity) \cliub{indicate that} components with identical convective velocity as determined by figure \ref{uc_nearWall_tripanel_localmean} also have nearly identical wall-normal coherence. This behavior, also suggested by \citet{delAlamo2009}, agrees with the hypothesis that a random arrangement of similar basic structures with dimensions given by $\lambda_z^+ = (\lambda_x^+)^{\frac{2}{3}}$, leads to the long-tailed behavior of the contours in figure \ref{uc_nearWall_tripanel_localmean}.

\begin{figure}
\centering
\includegraphics[scale=0.33]{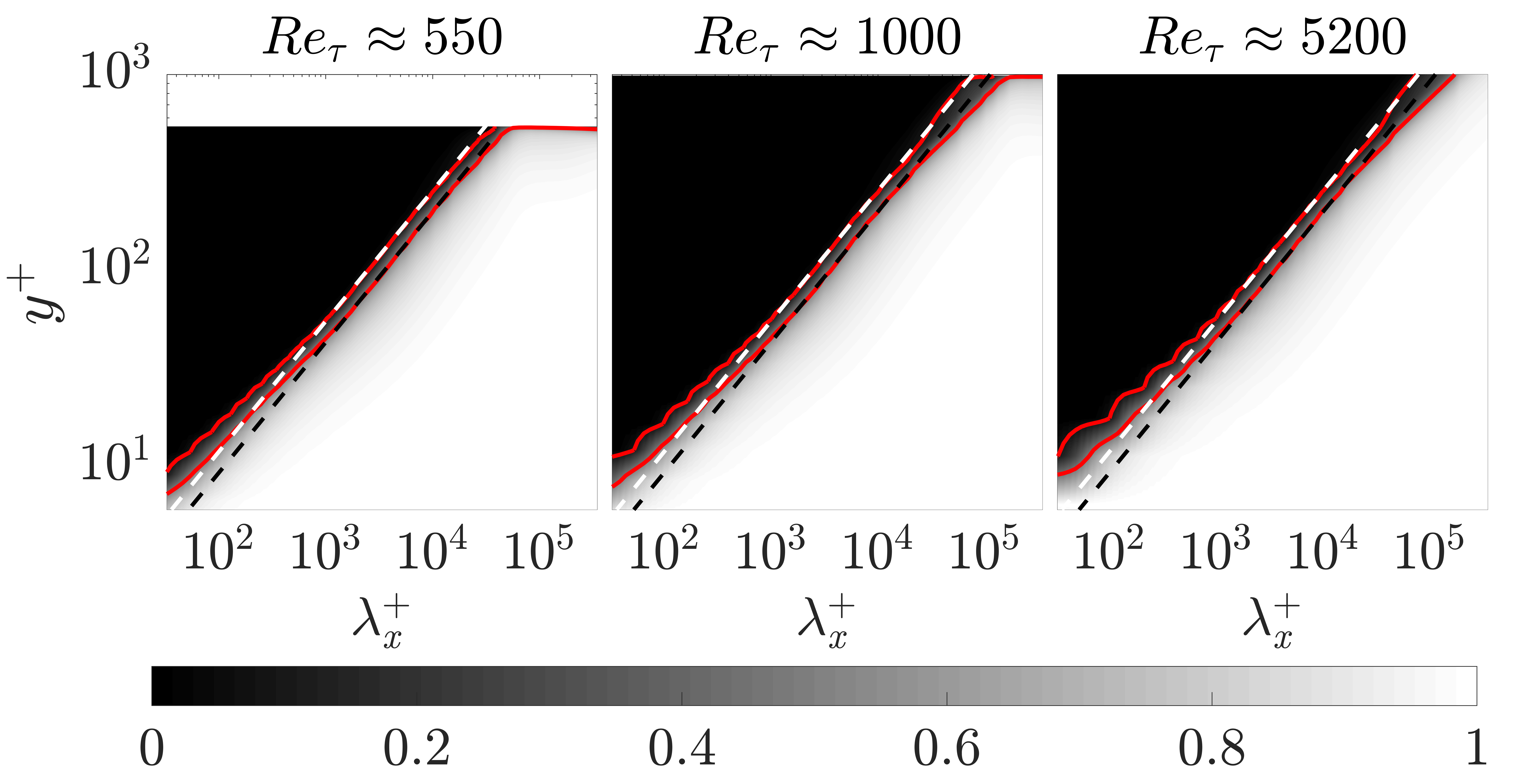} 
\caption{Two-point spectral coherence of streamwise velocity fluctuations $\chi_{u'; y'y}$ at $Re_{\tau}\approx550$, $Re_{\tau}\approx1000$, and $Re_{\tau}\approx5200$, as defined in (\ref{SpecCohdefn}), between ${y'}^+ \approx y_{\text{sublayer}}^+ = 5$ and wall-normal locations above it for wavelengths defined by $\lambda_z^+ = (\lambda_x^+)^{\frac{2}{3}}$ and phase speeds indicated by the dashed maxima line in figure \ref{uc_phi_monopanel}. 
The solid red lines serve to indicate the boundaries of regions of high/low coherence and are isocontours of spectral coherence with values 0.1 and 0.5. The white dashed lines are $y^+=0.55(\lambda_x^+)^{\frac{2}{3}}$ and the black dashed lines are $y^+=0.43(\lambda_x^+)^{\frac{2}{3}}$ and they serve as fits to the red lines.
Perfectly coherent signals have a spectral coherence of $1$, and incoherent signals have a spectral coherence of $0$.}
\label{1D_coherence_uvw_sqrtlambdaz}
\end{figure}

\cliua{Figure \ref{1D_coherence_uvw_sqrtlambdaz}} shows the spectral coherence with respect to the viscous sublayer location ($y_{\text{sublayer}}^+ \equiv 5$) $\chi_{u';y_{\text{sublayer}}y}(k_x,k_z,c)$ along $\lambda_z^+ = (\lambda_x^+)^{\frac{2}{3}}$. 
The monotonic behavior of the spectral coherence in figure \ref{1D_coherence_uvw_sqrtlambdaz} implies that structures larger in $(x,z)$ are also larger in $y$. The wall-normal coherence for $\lambda_x^+ \gtrsim 200$ \cliub{indicated by the dashed lines overlain on the (red) solid contours representing spectral coherences of $0.1$ and $0.5$} shows \cliub{an aspect ratio} $y^+\sim(\lambda_x^+)^{\frac{2}{3}}$. The minimum wall-normal coherence length associated with these larger wavelengths is $\sim 15$ wall units, which is the approximate location of the buffer layer and also the location of the well-known peak in the root-mean-square (RMS) streamwise velocity fluctuations; see for example, \citet{Lee2015}. This self-similarity \cliua{represented} by \cliua{a} power-law relationship at larger wavelengths is \cliub{also} suggestive of the attached-eddy structures proposed by \citet{Townsend1976,perry1986theoretical,marusic2019attached}.

From the power-law behavior $y^+\sim(\lambda_x^+)^{\frac{2}{3}}$ for $\lambda_x^+ \gtrsim 200$ shown in figure \ref{1D_coherence_uvw_sqrtlambdaz}, we \cliua{can also extract} the structure inclination angle contributing to this self-similar behavior.
The $0.1$ and $0.5$ spectral coherence contour is fitted by $y^+ = \alpha (\lambda_x^+)^{\frac{2}{3}}$ with $\alpha=0.55$ (white dash lines) and $\alpha=0.43$ (black dash lines), respectively. \cliuf{
We select the spectral coherence contours as 0.1 and 0.5 to fit the scaling laws because we observe significant variation of coherence between this range in figure \ref{1D_coherence_uvw_sqrtlambdaz}, while outside of this range, the coherence show saturation. Such a saturation phenomenon is also observed in the coherence computed from the experimental data; see e.g., figure 4 of \citet{baars2016spectral} and figure 5(b) of \citet{baars2017self}. Furthermore, the contours of two-dimensional spectral coherence of $0.1$, $0.3$, and $0.5$ are shown to collapse when scaled with the wall-normal height of the structures; indicating the presence of self-similar structures, see figure 4 of \citet{madhusudanan2019coherent}. 
}

\begin{figure}
    \centering
    \includegraphics[width=3in]{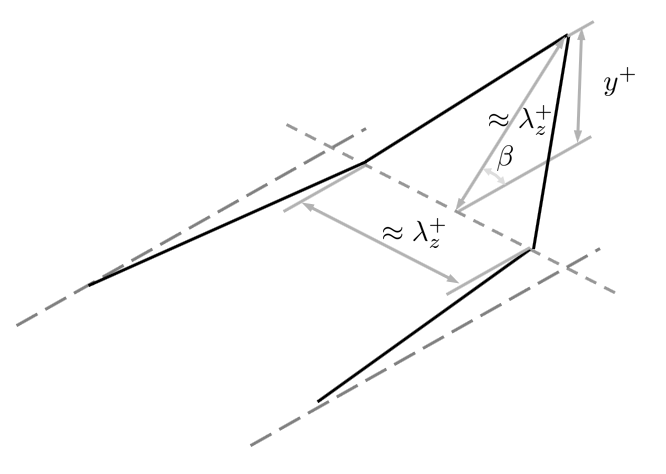}
    \caption{\cliuf{The structures of height/width aspect ratio 1 with an inclination angle $\beta$ in analogy with \citet{perry1982mechanism}.}}
    \label{fig:attached_eddy}
\end{figure}

Thus, $y^+ \sim \lambda_z^+$ with a constant of proportionality \cliuf{$\alpha$} between $0.55$ and $0.43$, respectively, imply that the projection of the structures onto the cross-stream plane has a smaller height than width. If we assume the structures contributing to the spectral coherence \cliuf{have a height/width aspect ratio of $1$ as depicted in the cartoon in figure \ref{fig:attached_eddy}}, then the dimensions of the cross-plane projection of the structure represented by spectral coherence between $0.1$ and $0.5$ implies a tilt angle \cliuf{$\beta=\text{arcsin}(\alpha)$} between approximately $25^\circ$ and $33^\circ$. 

\citet{Townsend1976} suggests an inclination angle of $\sim 30^\circ$ \cliue{for} attached double roller eddies (a pair of counter-rotating, inclined, approximately streamwise vortex structures) to explain the experimental observations of \citet{Grant1958}. Experimental observations in turbulent boundary layers, on the other hand, have yielded inclination angles between $15^\circ$ and $20^\circ$ \citep{Brown1977,Marusic2007,Carper2004}. The latter inclination angles were calculated using two-point temporal correlations and Taylor's hypothesis. In turbulent channel flows, hairpin vortices have been \cliue{the} focus of considerable interest. Although a single well-defined inclination angle cannot be associated with hairpin vortices, the inclinations of hairpin like structures vary from $12^\circ$ (elongated legs) to $45^\circ$ (hairpin heads) \citep{Adrian2007}. The present work does not restrict the structures contributing to the convective velocities to any one of the structures discussed above, but does provide an inclination angle, assuming structures are roughly of aspect ratio $1$, which is within the range of previous observations.

In figure \ref{1D_coherence_uvw_sqrtlambdaz}, the structure inclination angle predicted by this model also shows Reynolds number invariance, which is consistent with experimental observations \citep{Marusic2007}. For different Reynolds numbers, $y^+ = \alpha (\lambda_x^+)^{\frac{2}{3}}$ with $\alpha=0.55$ and $\alpha=0.43$, corresponding to tilt angles $25^\circ$ and $33^\circ$, always give good approximations for $0.1$ and $0.5$ spectral coherence, respectively. Structure inclination angles inferred from the cross correlation \cliue{of $x$ in the experimental studies of \citet{Marusic2007}} are found to be invariant over 3 orders of magnitude change in Reynolds number. 

Our results reveal that the contributions from the relatively larger scale structures lead to the elevated velocities in the viscous sublayer seen in figure \ref{uc1D} and that these structures have dimensions given by $y^+ \sim \lambda_z^+ \sim (\lambda_x^+)^{\frac{2}{3}}$ with a minimum size $\approx 15$ friction units, which is the approximate location of the buffer layer. The inclination angles of these structures do not vary \cliua{with} Reynolds number, which is consistent with experimental observation\cliua{s}. 
\cliue{These observations are} consistent with the attached-eddy hypothesis in that they predict wall-attached structures that are self-similar in the cross-plane and contribute to the dominant convection. However, the attached-eddy hypothesis predicts that these structures are also self-similar in the streamwise direction, which \cliua{our \cliuf{approach} does not show}.

\section{Term-by-term analysis of scale-dependent convective velocities}
\label{sec:term}

\cliua{We next use the \cliue{input-output} framework to analyze the \cliub{contribution} of different linear mechanisms to the \cliue{scale-dependent} convective velocity \cliue{of the streamwise velocity fluctuations}.}

We follow the method \cliuf{shown in equation (2.11)} of \citet{delAlamo2009} \cliue{to obtain the normalized deviation of the convective velocity from the mean velocity \cliuf{contributed from various terms}. In particular, we multiply} the $x$-momentum in equation (\ref{NSPFT2}) by $\hat{u}'^*$ and take the imaginery part of the result to obtain:
\begin{align}
    \frac{u_c-\bar{u}(y)}{\bar{u}(y)}=\frac{ \overbrace{k_x \Real{}\Big\{ \langle \hat{p}' \hat{u}'^*\rangle\Big\}}^{\text{IIa}}+\Imag{}\bigg\{\overbrace{ \frac{d\bar{u}}{dy} \langle\hat{v}' \hat{u}'^*\rangle}^{\text{IIb}}  \overbrace{-\frac{1}{Re_\tau} \langle\hat{u}'^*  \p^2_{yy}\hat{u}'\rangle}^{\text{IIc}}\overbrace{-\langle\hat{f_u}' \hat{u}'^*\rangle}^{\text{III}}\bigg\}}{k_x\bar{u}(y) \langle \hat{u}' \hat{u}'^*\rangle}. 
    \label{delAlamoTerm}
\end{align}
\cliub{Here $\text{Re}\{\cdot\}$ and $\text{Im}\{\cdot\}$ represent the respective real part and imaginary part of the argument.} \cliub{The terms in equation (\ref{delAlamoTerm}) represent the relative contributions of} the pressure term (IIa), the mean shear term (IIb), and the viscous term (IIc), each normalized by $k_x \cliu{\bar{u}(y)}\langle \hat{u}' \hat{u}'^*\rangle$.

We compute each term in (\ref{delAlamoTerm}) by modifying the output operator in (\ref{IOmap}) and then computing the cross-spectra through an appropriate modification of (\ref{phipsi1}). For example, we can \cliub{use the output operator \cliue{corresponding to the fluctuating pressure }in (\ref{output_vorticity_pressure_Coeffs}) to obtain $\mathcal{G}_{\hat{p}'}$ and then compute the cross-spectra as}\cliua{
\begin{align}
\langle\hat{p}'\hat{u}'^*\rangle
&
=   \mathcal{G}_{\hat{p}'} \langle\bld{\hat{f}}'  \bld{\hat{f}}'^*\rangle \mathcal{G}_{\hat{u}'}^* 
=   \mathcal{G}_{\hat{p}'}  \mathcal{G}_{\hat{u}'}^*.
\end{align}}The other terms in (\ref{delAlamoTerm}) can be computed in a similar manner.

\begin{figure}
\centering 
\begin{subfigure}[b]{\textwidth}
	
 \large{ (a) }
 
\centering
\includegraphics[scale=0.29]{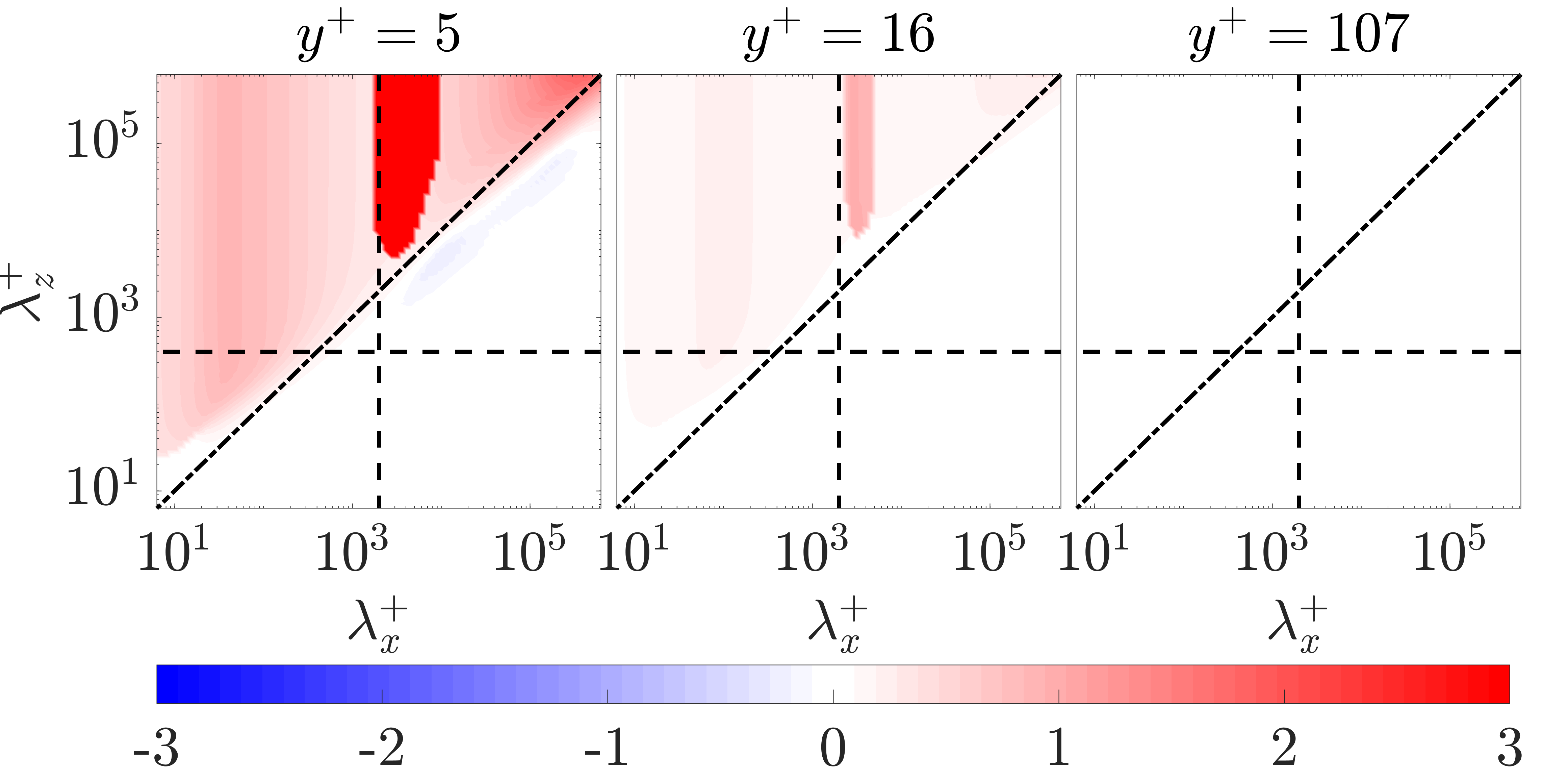}

\end{subfigure}

\begin{subfigure}[b]{\textwidth}
	
\large{ (b) }

\centering
\includegraphics[scale=0.29]{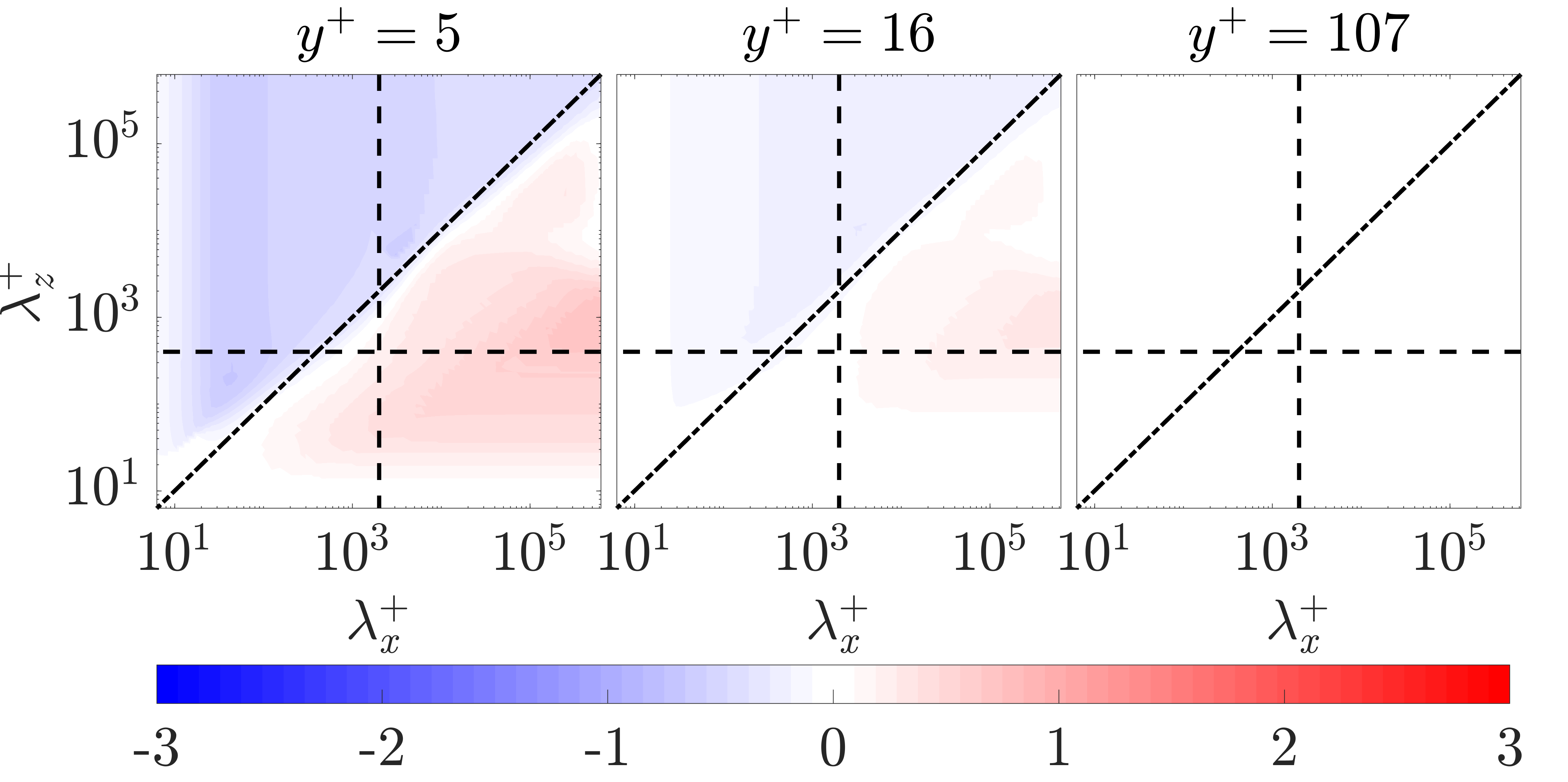}

\end{subfigure}

\begin{subfigure}[b]{\textwidth} 
		\large{ (c) }
		
		\centering
		\includegraphics[scale=0.29]{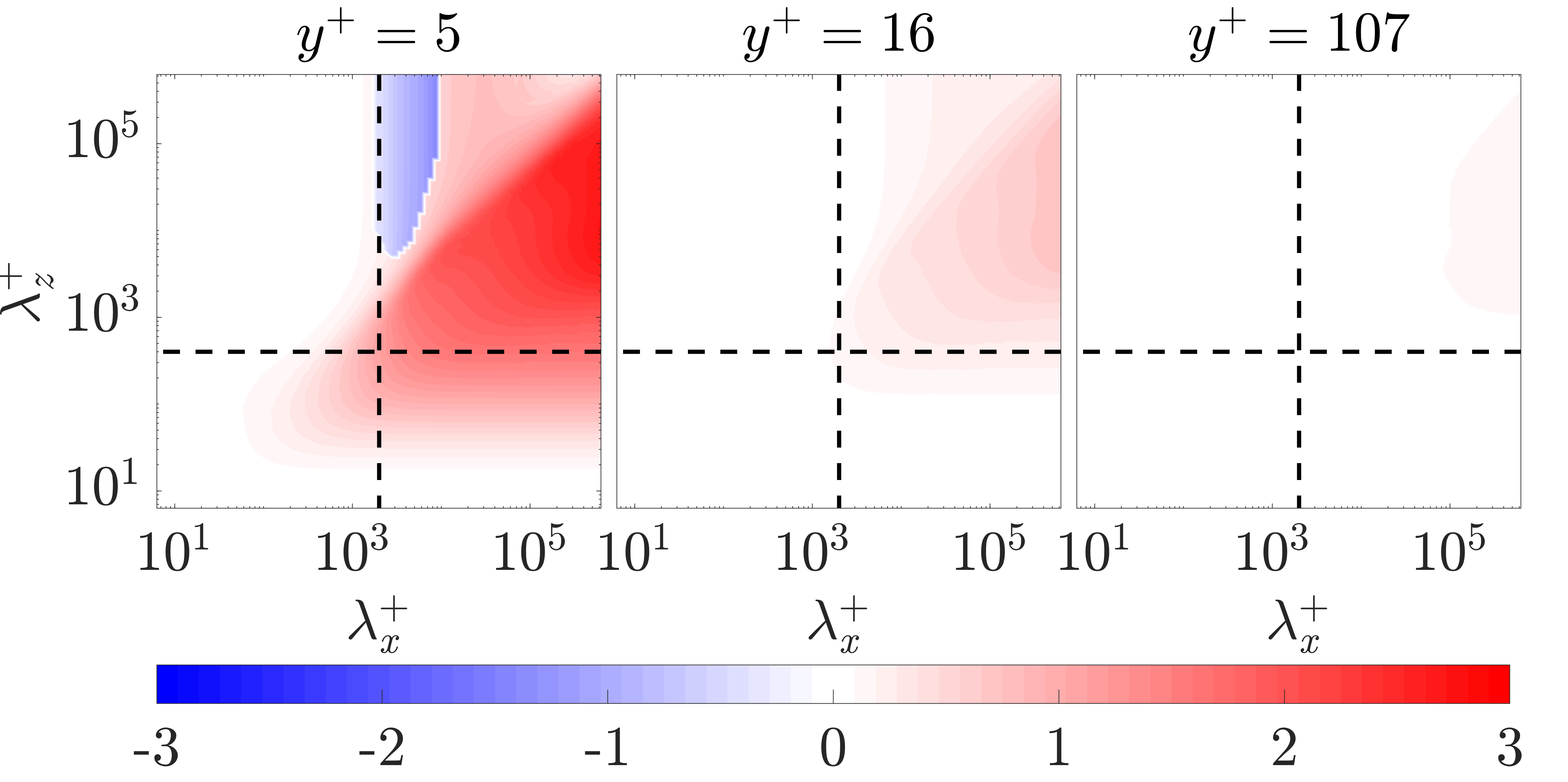}

\end{subfigure} 
\caption{Different linear terms' contributions to scale-dependent convective velocities $(u_c(y;k_x,k_z)-\bar{u}(y))/\bar{u}(y)$ quantified using equation (\ref{delAlamoTerm}): (a) the pressure term (IIa), (b) the mean shear term (IIb), and 
(c) the viscous term (IIc). All \cliue{terms} are normalized by $k_x \bar{u}(y)\langle \hat{u}' \hat{u}'^*\rangle$. The Reynolds number is $Re_{\tau}=1000$.
The black dashed lines are given by $(\lambda_x,\lambda_z) = (2,0.4)$, and the black dash-dot lines in (a) and (b) are $\lambda_z^+ = \lambda_x^+$. \cliuf{Note: the white region of the color map represents a value close to zero.} }
	\label{DiffTerm2Conv}
\end{figure}

\cliua{Figure \ref{DiffTerm2Conv}(a), (b), and (c) show the respective contributions from the pressure term (IIa), the mean shear term (IIb), and the viscous term (IIc) to the scale-dependent convective velocity of the streamwise velocity fluctuations ($Re_{\tau}\approx1000$) \cliue{at the same three wall-normal locations as in figure \ref{uc_nearWall_tripanel_localmean}}.}
\cliuf{As shown in figure \ref{DiffTerm2Conv}(a), the} pressure plays an important role for the intermediate \cliue{scale} structures ($\lambda_x\approx2$ and $\lambda_z>\lambda_x$), which \cliub{supports our conjecture that the \cliuf{discontinuity} in these scales shown in figure \ref{uc_nearWall_tripanel_localmean} is related to the pressure.} \cliuf{\cites{Luhar2014} figure 12(a) also showed a discontinuity of the scale-dependent convective velocity of wall-pressure computed using resolvent analysis and the maximum of the PSD to define the convective velocity.} \cliuf{As discussed in Section 5, using the center of gravity of the PSD to define the convective velocity eliminates the discontinuity. A similar smoothing effect resulting from the use of the center of gravity definition versus the maximum value of the PSD was also observed in figures 12(a) and (b) of \citet{Luhar2014}, where the authors compared these two convective velocity definitions for pressure fluctuations.} \cliub{The overall convective velocity of these intermediate-scale structures also} includes contributions from both the mean shear (figure \ref{DiffTerm2Conv}(b)) and the viscous terms (figure \ref{DiffTerm2Conv}(c)), which indicates that multiple physical mechanisms are at play.

\cliua{For the large-scale structures with $(\lambda_x,\lambda_z) > (2,0.4)$, the deviation of convective velocity from the mean} is \cliub{primarily due to the} viscous and the mean shear terms. \cliua{In the viscous sublayer ($y^+\approx 5$), the viscous term provides a relatively larger contribution to the deviation of the convective velocity from the mean than the mean shear term, whereas these two terms provide approximately equal contribution to the convective velocity \cliue{in} the buffer layer ($y^+\approx 15$). 

}

\cliua{For structures with small \cliue{streamwise and spanwise} wavelengths}; i.e., $\lambda_x^+\lesssim 10$ and $\lambda_z^+\lesssim 10$, all of the terms in (\ref{delAlamoTerm}) are negligible \cliuf{(as indicated by the white region of the colormap).} \cliuf{This suggests that they convect at the local mean velocity or that their convective velocity is not captured through the linear terms retained in our approach.} However, \cliub{as previously noted, the} nonlinear fluctuation-fluctuation interactions \cliub{likely dominate at these scales, so linear analysis is unlikely to fully explain the mechanisms at play. Understanding the effects of nonlinearity is beyond the scope of the current paper, so we leave this as a topic of future work.}

\begin{figure}
\centering 
\begin{subfigure}[b]{\textwidth}
	
 \large{ (a) }
 
\centering

\includegraphics[scale=0.31]{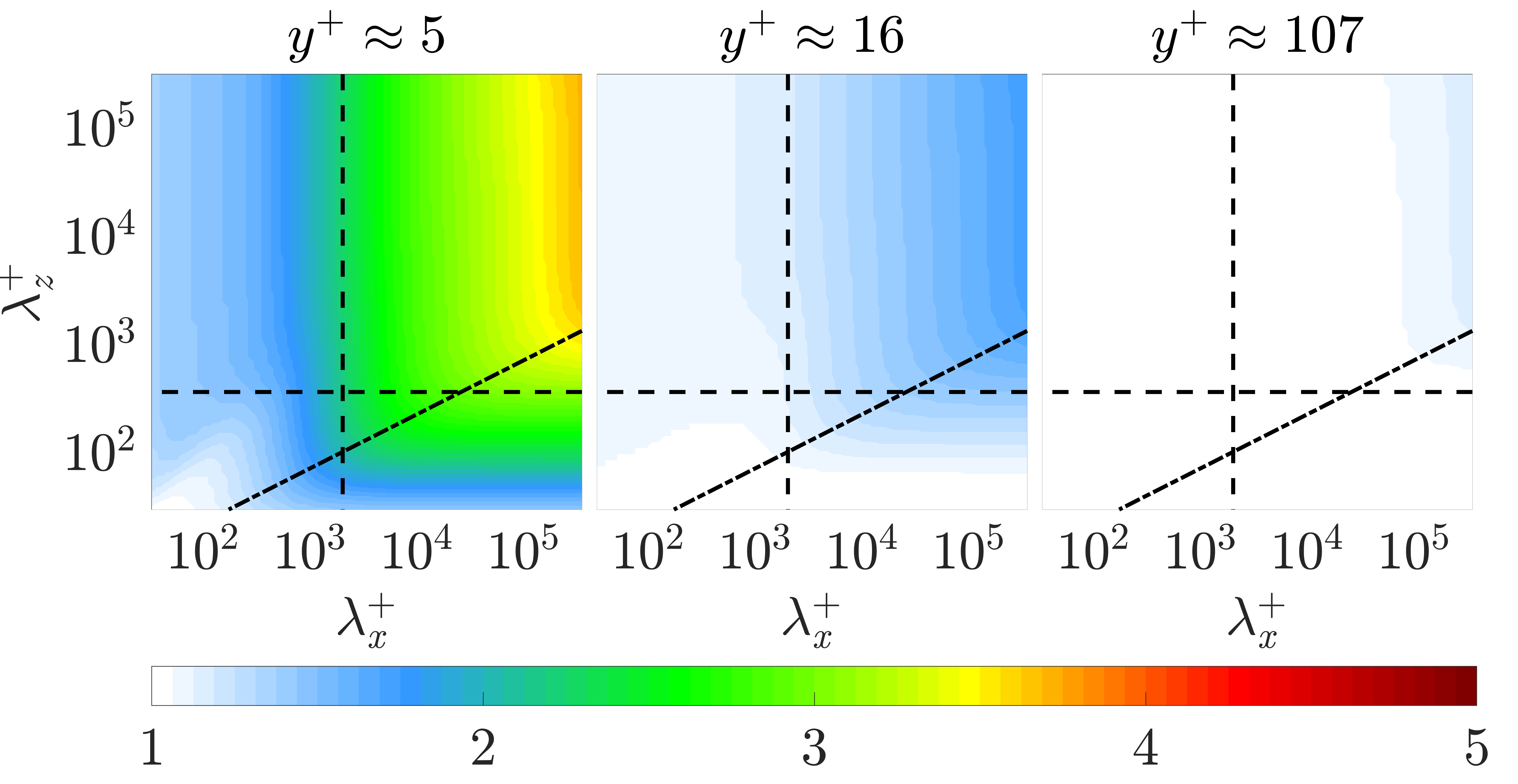}

\end{subfigure}

\begin{subfigure}[b]{\textwidth}
	
\large{ (b) }

\centering

\includegraphics[scale=0.31]{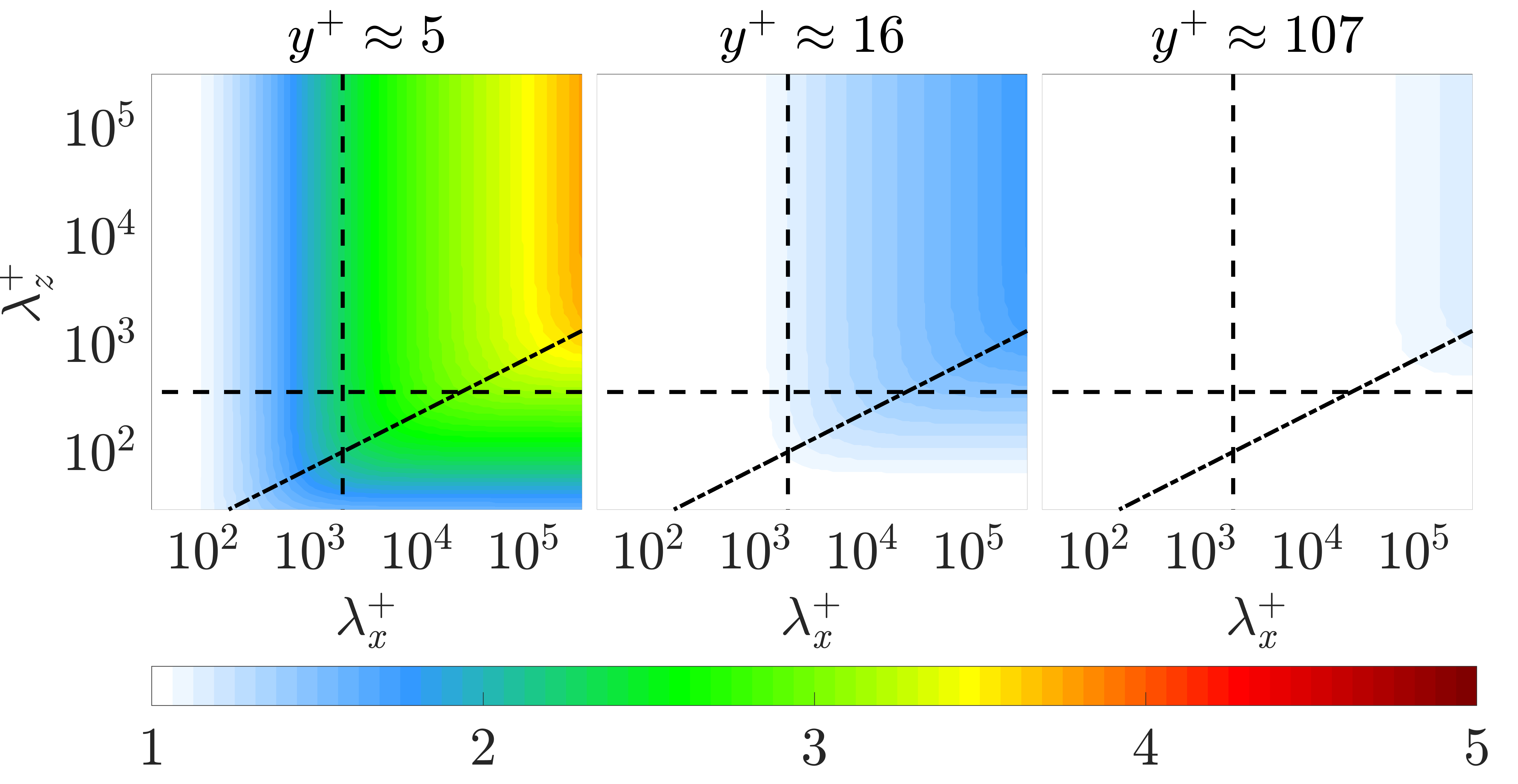}
\end{subfigure}

\begin{subfigure}[b]{\textwidth} 
		\large{ (c) }
		
		\centering
		
		\includegraphics[scale=0.31]{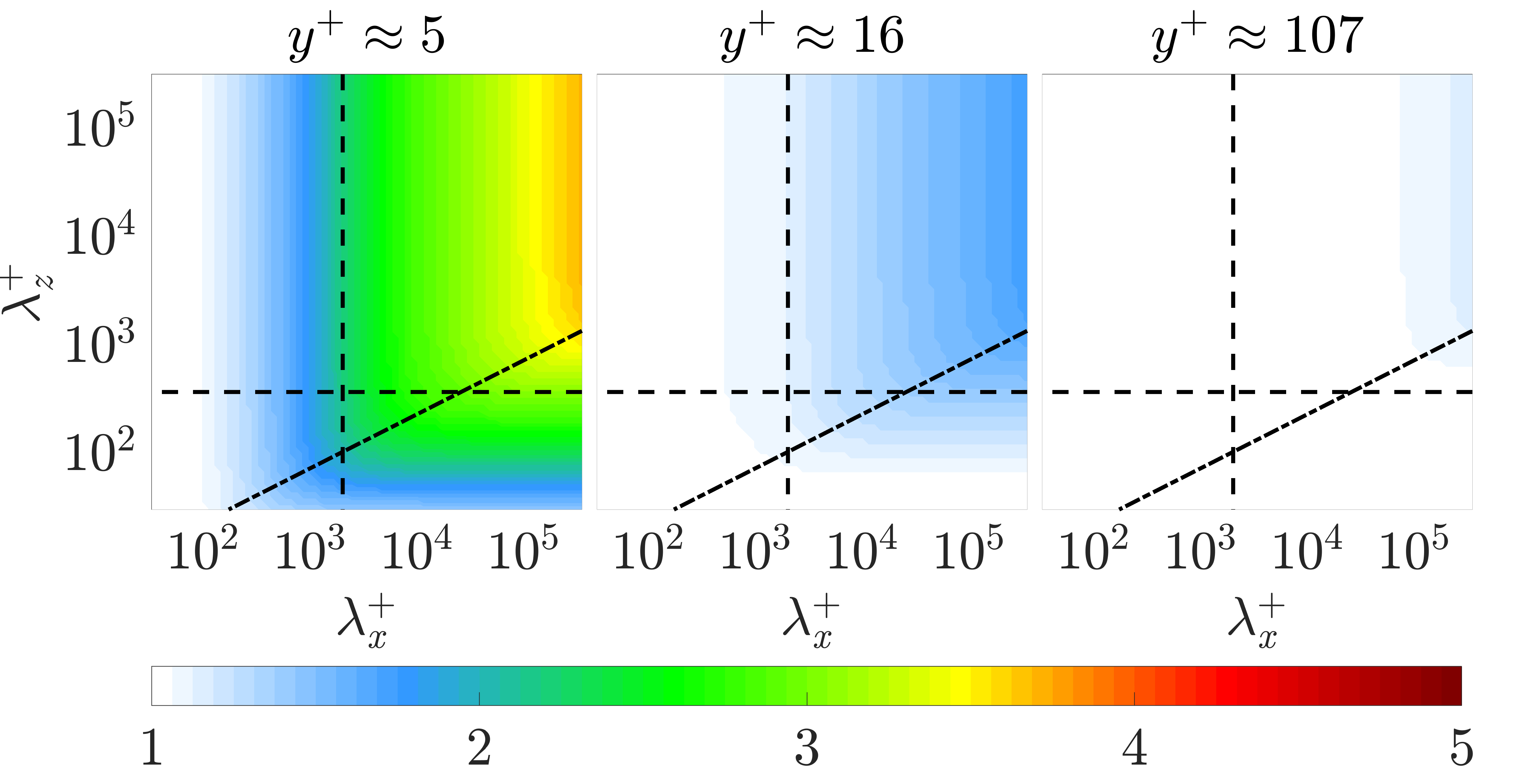}
		
\end{subfigure} 
\caption{ $\cliu{u}_c(y,\lambda_x,\lambda_z)/\bar{u}(y)$ at $\Rey_{\tau}=1000$ (a) neglecting the mean shear term \cliub{as (\ref{sysopernoshear})}, (b) neglecting the coupling from the pressure and mass conservation \cliub{as (\ref{sysopernopre})}, and
(c) neglecting the mean shear term, the pressure term and mass conservation together \cliub{as (\ref{sysoperOnlyVis})}. 
The black dashed lines are given by $(\lambda_x,\lambda_z) = (2,0.4)$.
The black dash-dot lines are $\lambda_z^+ = \frac{5}{2}\sqrt{\lambda_x^+}$, which fit through the knee of these contours.}
	\label{Fig:convNoshear}
\end{figure} 

To gain more insight into \cliub{the effect of each term}, we next compute the convective velocities \cliub{by neglecting the contribution of different terms in the linear dynamics that form the spatio-temporal transfer function in (\ref{IOmap}). In each case, we first describe how neglecting the term(s) of interest alters these operators and then evaluate the effect on the convective velocity.} \cliub{Setting} the mean shear term \cliub{to zero reduces the linear operator in (\ref{sysOpsDefn})} to
\begin{align}
\mathcal{L} &:=
\begin{bmatrix}
(\im k_x (\bar{u} - c)  
- \frac{1}{Re_\tau}\hat{\Delta})\mathbf{I}_{3\times 3} & \hat{\nabla} \\
\hat{\nabla}^\mathsf{T} & 0 
\end{bmatrix}. 
\qquad
\label{sysopernoshear}
\end{align}
\cliub{In this case, the operators $\mathcal{B}$ and $\mathcal{C}$ in (\ref{output_velocity_Coeffs}) remain the same.} \cliub{We note that although} the mean shear term is zero, $\bar{u}(y)$ is still a function of wall-normal location; therefore there is still shear imposed by the mean flow. Figure \ref{Fig:convNoshear}(a) \cliue{shows that the} convective velocities of the large scales continue to deviate from the mean velocity even when we eliminate the linear term associated with the mean shear. However, \cliub{the knee occurring at} $\lambda_z^+=\frac{5}{2}\sqrt{\lambda_x^+}$ shown in figure \ref{Fig:convNoshear}(a) is different from \cliub{that at} $\lambda_z^+={\lambda_x^+}^\frac{2}{3}$ in figure \ref{uc_nearWall_tripanel_localmean} \cliub{based on the full linear \cliuf{approach}}. This is consistent with figure \ref{DiffTerm2Conv}(b), which indicates that the mean shear term plays a role in the self-similarity predicted in this \cliuf{approach}.

We \cliub{next isolate} the role of the pressure. For this analysis we \cliub{group the effect of} the pressure gradient and the mass conservation \cliub{terms \cliue{because} they both contribute to the nonlocality of the turbulent flow.} \cliub{This relationship can be understood by viewing} the pressure in the momentum equation as the Lagrange multiplier that enforces the divergence-free velocity field; see e.g., section 5.6.2 in \citet{Schmid2001Stability}. Neglecting \cliub{both} the pressure term and \cliue{the divergence free constraint reduces the operators in the} input-output map $\mathcal{G}_{\hat{u}'}=\mathcal{C}_{\hat{u}'}\mathcal{L}^{-1}\mathcal{B}$ to:
\begin{align}
\mathcal{C}_{\hat{u}^\prime}:= \begin{bmatrix} 1 & \bld{0}_{1\times 2}\end{bmatrix},
\mathcal{L}:=\begin{bmatrix}
(\im k_x (\bar{u} - c)  
- \frac{1}{Re_\tau}\hat{\Delta})\mathbf{I}_{3\times 3} +   \dd{\bar{u} }{y} \mathbf{S}
\end{bmatrix}, 
\mathcal{B} :=\mathbf{I}_{3\times 3}
. 
\label{sysopernopre}
\end{align}

The resulting convective velocities in figure \ref{Fig:convNoshear}(b) are similar to those in figure \ref{Fig:convNoshear}(a) with \cliue{the} mean shear \cliue{term} set to zero. Neither of these terms appear to be responsible for the influence of the large-scale structures that leads to the observed behavior of the convective velocity in the near-wall region.  They also do not reproduce the $\lambda_z^+={\lambda_x^+}^{\frac{2}{3}}$ scaling, but they do emit self-similar structures with a different scaling exponent, $\lambda_z^+=\frac{5}{2}{\lambda_x^+}^{\frac{1}{2}}$. 

\begin{figure}

\hspace{2.4cm} \large{ (a) }\hspace{3.3cm}
\large{ (b) }\hspace{3.3cm}
\large{ (c) }

		\centering
		
		\includegraphics[scale=0.31]{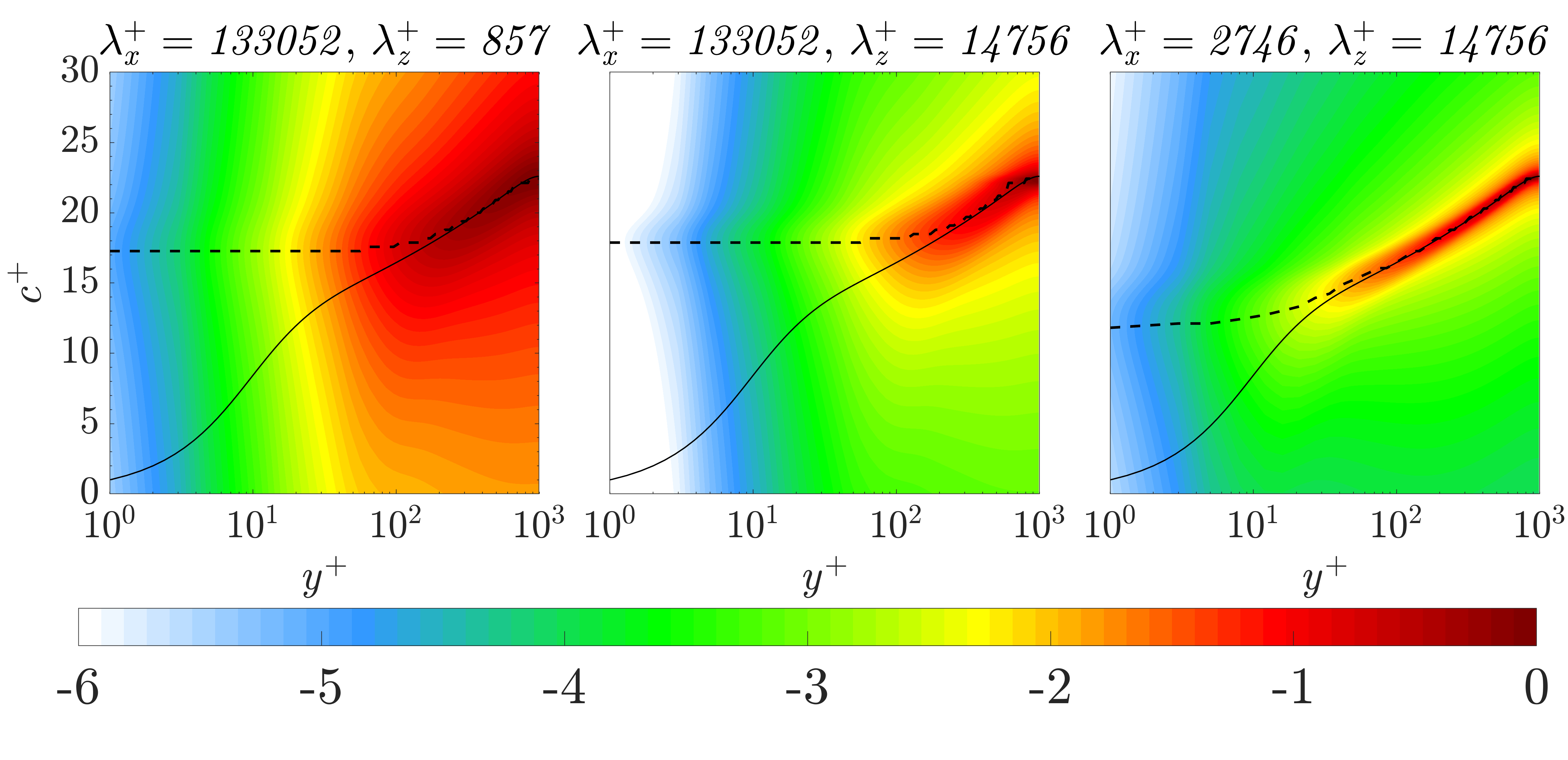}
 
	\caption{Power spectral density of streamwise velocity fluctuations over wall-normal location $y^+$ and phase speed $c$ $\frac{\Phi_{\hat{u}'}(y;k_x,k_z,c)}{\max\limits_{c,y}(\Phi_{\hat{u}'}(y;k_x,k_z,c))}$ at $Re_\tau=1000$ from model (\ref{sysoperOnlyVis}) \cliue{for representative} large-scale structures $\vartriangleleft (\lambda_x^+,\lambda_z^+)=(133052,857)$, $\vartriangleright (\lambda_x^+,\lambda_z^+)\approx(133052,14756)$, and intermediate-scale structures $\triangle \; (\lambda_x^+,\lambda_z^+)\approx(2746,14756)$. The color is in base 10 logarithmic scale. The black solid lines represent the mean streamwise velocity profile, and the black dashed lines are convective velocites, which are defined in (\ref{ucdefn}) as \cliue{the} phase speed that maximizes the PSD of the streamwise fluctuations $\Phi_{\hat{u}'}(y;k_x,k_z,c)$.}
	\label{energyVisOnly}
\end{figure}

In order to evaluate their combined effect, we next neglect the contributions of both the mean shear and pressure terms, leaving only the advective and viscous terms. The resulting input-output \cliuf{based approach} for the streamwise velocity fluctuations \cliub{is given by} $\hat{u}'=\mathcal{C}_{\hat{u}^\prime}\mathcal{L}^{-1}\mathcal{B}\hat{f}_x$ with
\begin{align}
\mathcal{C}_{\hat{u}^\prime}:= \begin{bmatrix} 1 \end{bmatrix},
\quad
\mathcal{L} :=
\begin{bmatrix}
\im k_x (\bar{u} - c)  
- \frac{1}{Re_\tau}\hat{\Delta}) 
\end{bmatrix}, 
\quad
\mathcal{B} :=
\begin{bmatrix}
1
\end{bmatrix}.
\label{sysoperOnlyVis}
\end{align} Figure \ref{Fig:convNoshear}(c) \cliue{shows the resulting convective velocity contours, which are similar to the results in panels (a) and (b)}. In particular, they reproduce the influence of the large-scale structures in the near-wall and buffer regions seen in the full LNS based \cliuf{approach}. Figure \ref{energyVisOnly} plots the \cliue{power spectral density} of the streamwise velocity fluctuations at different phase speeds $c^+$ and wall-normal locations $y^+$ computed using the model in equation (\ref{sysoperOnlyVis}). Although \cliue{there are} some difference\cliub{s} from the results obtained using the full LNS \cliue{system} shown in figure \ref{energy}, the phase speed that maximizes the energy spectrum; i.e., the convective velocity, still asymptotes to a constant value near the wall for large wavelength structures. 

The main difference between these results and the full LNS based \cliuf{approach} is that they show the same $\lambda_z^+=\frac{5}{2}{\lambda_x^+}^{\frac{1}{2}}$ scaling as the previous model in (\ref{sysoperOnlyVis}) with the influence of the pressure and mean shear removed. The inability to reproduce the correct aspect ratio for the self-similar structures suggest that their morphology is due to interactions between viscous mechanisms and other inviscid mechanism arising due to the interaction of the fluctuations with the mean shear $\text{d}\bar{u}/\text{d}y$ and the pressure, such as the lift-up effect \citep{Brandt2014} and the Orr mechanism \citep{Farrell1987,Jimenez2013}.  However, the prediction of the main trends and scale interactions suggest that this type of model may provide a good balance between accuracy and simplicity\cliuf{. We} next explore its potential as a viscous correction to Taylor's hypothesis.

\begin{figure}
    \centering
    \includegraphics[width=3in]{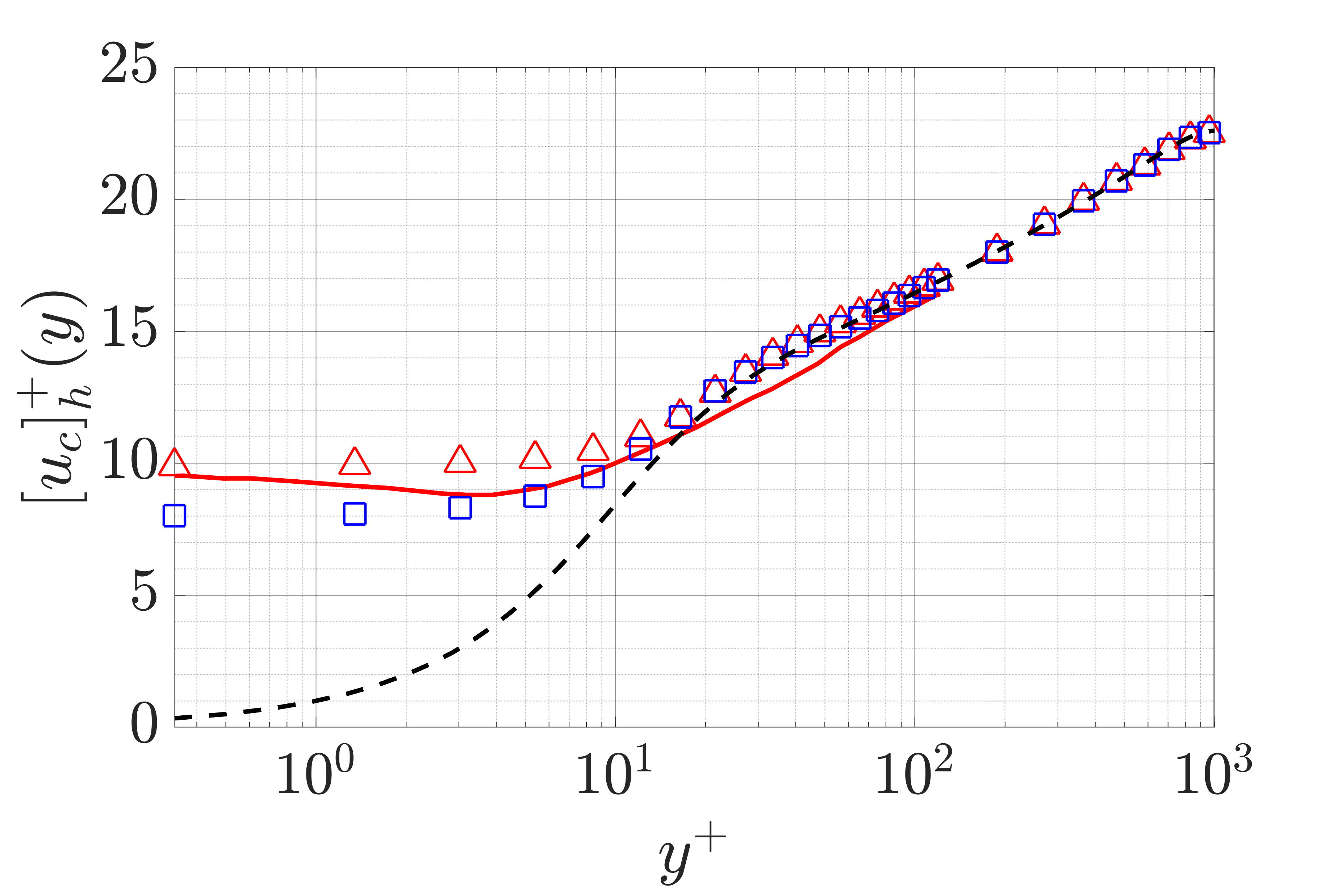}
    \caption{\cliuf{The average convective velocity of streamwise velocity fluctuations, 
	$[u_c]_h^+(y)$: ({\color{blue}\parbox{0.0825in}{$\hspace{-0.0070in}\vspace{-0.015in}\mathlarger{\square}$}}$\,$); computed using the viscous correction (7.6) with their corresponding weighting functions $h=\langle|\mathcal{F}_{xz}(u')|^2\rangle k_x^2$ and an averaging domain of $(\lambda_x^+,\lambda_z^+)>(500,80)$ at $Re_\tau=1000$. Results are plotted with convective velocities of streamwise velocity fluctuations computed from both the LNS based approach described herein for $Re_\tau=1000$: ({\color{red}\parbox{0.0825in}{$\hspace{-0.0070in}\vspace{-0.015in}\mathlarger{\triangle}$}}$\,$) and DNS data \citep{Geng2015} at $Re_\tau=932$: (\parbox{0.115in}{\color{Red} $\mline\mline$}). The \cliuf{black} dashed line is the turbulent mean velocity profile at $Re_{\tau}\approx1000$ from \citet{Lee2015}.}}
    \label{fig:viscous_average}
\end{figure}

We obtain this correction by rewriting equation (\ref{sysoperOnlyVis}) as
\begin{equation}
    \text{i}k_x(\bar{u}(y)-c)\hat{u}'-\frac{\hat{\Delta}}{Re_{\tau}}\hat{u}'=\hat{f}_x.
    \label{viscousonly}
\end{equation}\cliuf{
Figure \ref{fig:viscous_average} compares the average convective velocity of streamwise velocity fluctuations computed using the viscous correction (\ref{viscousonly}) with its corresponding weighting functions $h=\langle|\mathcal{F}_{xz}(u')|^2\rangle k_x^2$ and an averaging domain of $(\lambda_x^+,\lambda_z^+)>(500,80)$ at $Re_\tau=1000$ to the results from the full LNS based approach and convective velocities obtained from DNS data at $Re_\tau=932$ from \cite{Geng2015}. This figure shows that the average convective velocity predicted from the viscous correction shows excellent agreement with results obtained from DNS data for $y^+\in [5,15]$, but begins to deviate for $y^+\lesssim 3$. We therefore conclude that this viscous correction provides a potential dynamical modification on Taylor's hypothesis to improve the convective velocity estimates for use with experimental data.

This viscous correction introduced in equation (\ref{viscousonly}) could be augmented using an eddy viscosity, in the spirit of the eddy viscosity enhanced LNS equations introduced in \cite{reynolds1972mechanics}. Such a dynamical correction was previously shown to provide similar improvements in model-fidelity for certain structures as the inclusion of colored-in-time forcing~\citep{Zare2017}. This type model enhancement may be particularly relevant in this context because the pertinent terms would all be retained in the associated modification of the viscous correction proposed in equation (\ref{viscousonly}). Assessing the potential benefits of such an approach is a topic of future work.

}

\begin{figure}
\centering 
\begin{subfigure}[b]{\textwidth}
	
 \large{ (a) }
 
\centering

\includegraphics[scale=0.31]{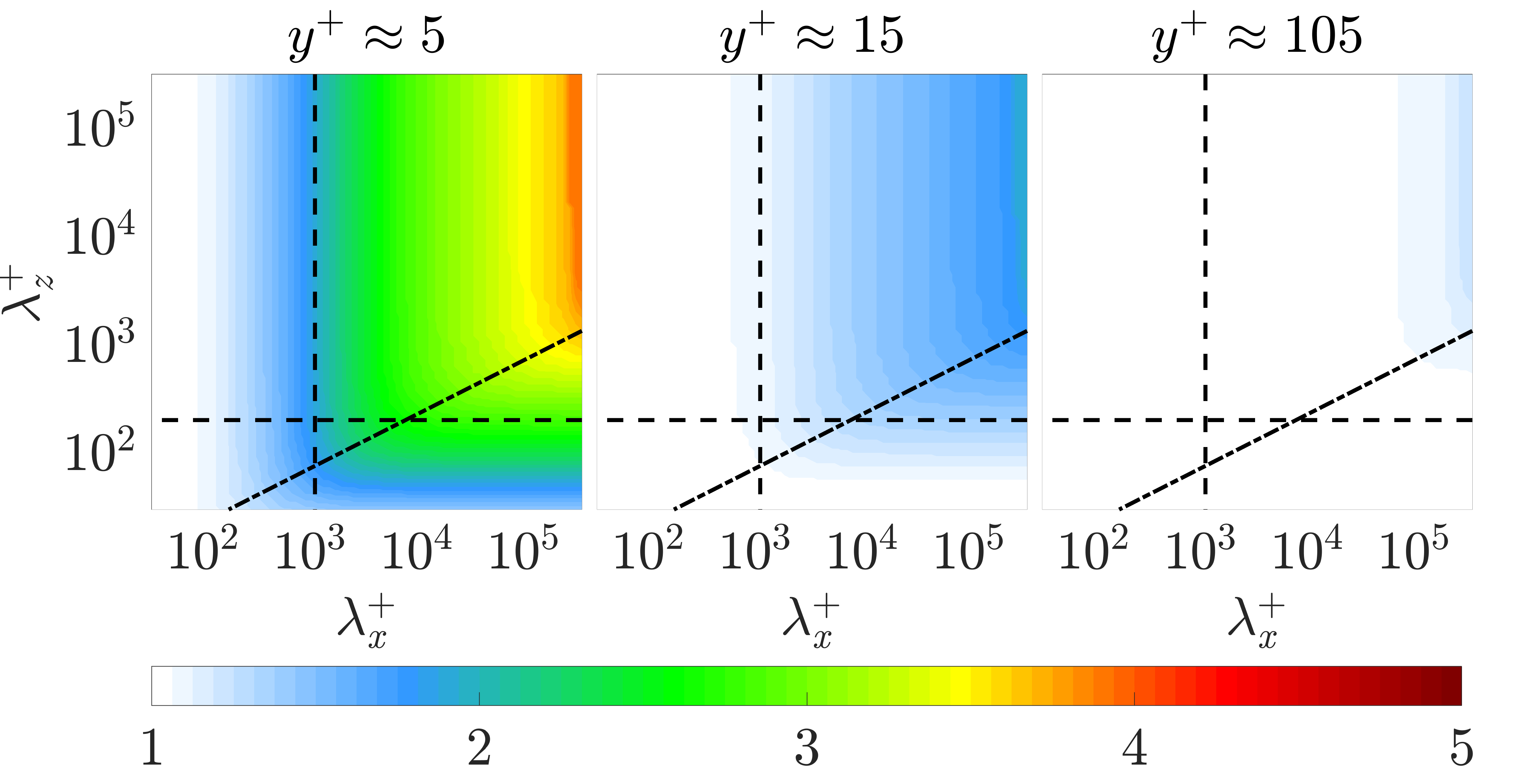}

\end{subfigure}

\begin{subfigure}[b]{\textwidth}
	
    \large{ (b) }

\centering

\includegraphics[scale=0.31]{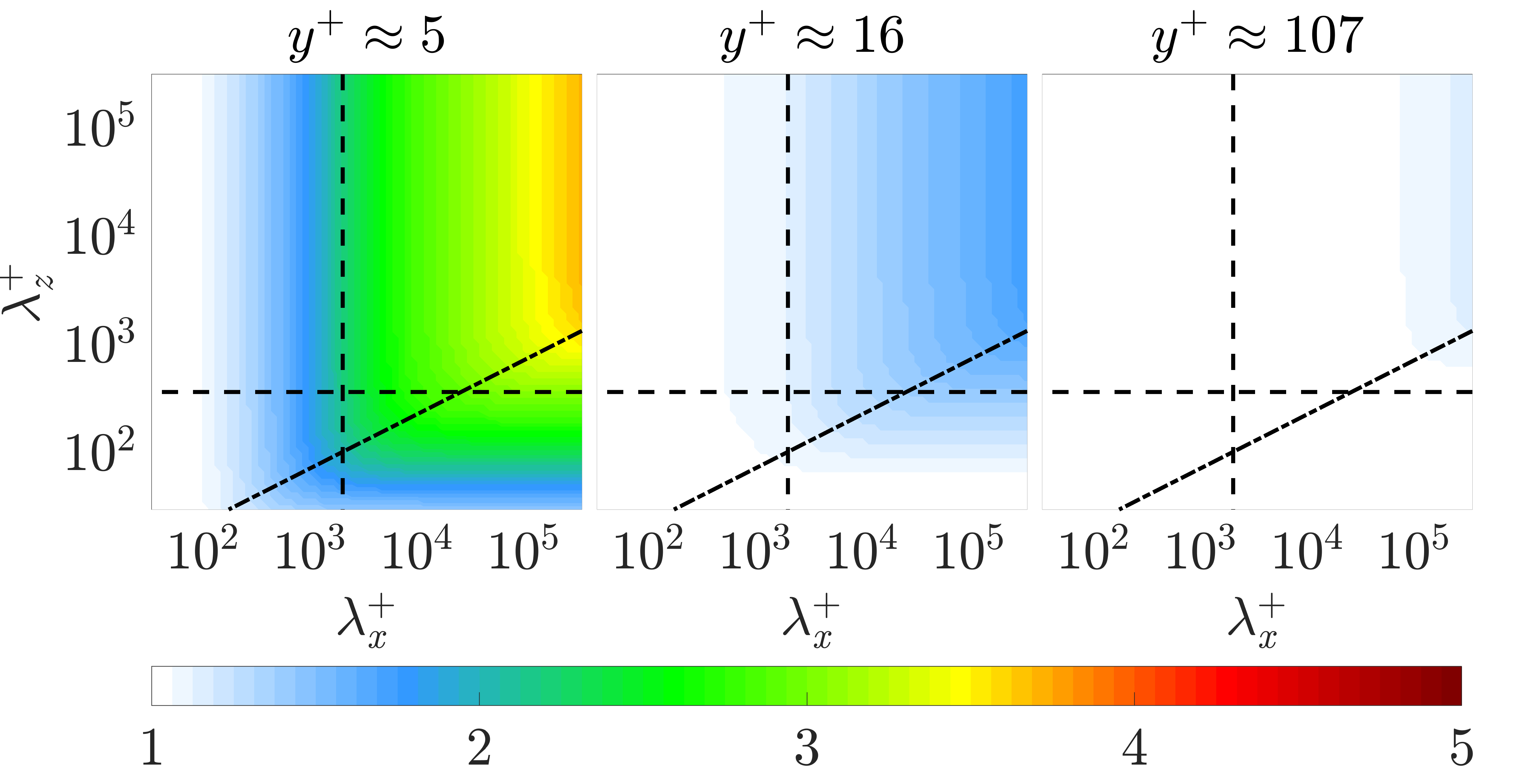}
\end{subfigure}

\begin{subfigure}[b]{\textwidth} 
		\large{ (c) }
		
		\centering
		
		\includegraphics[scale=0.31]{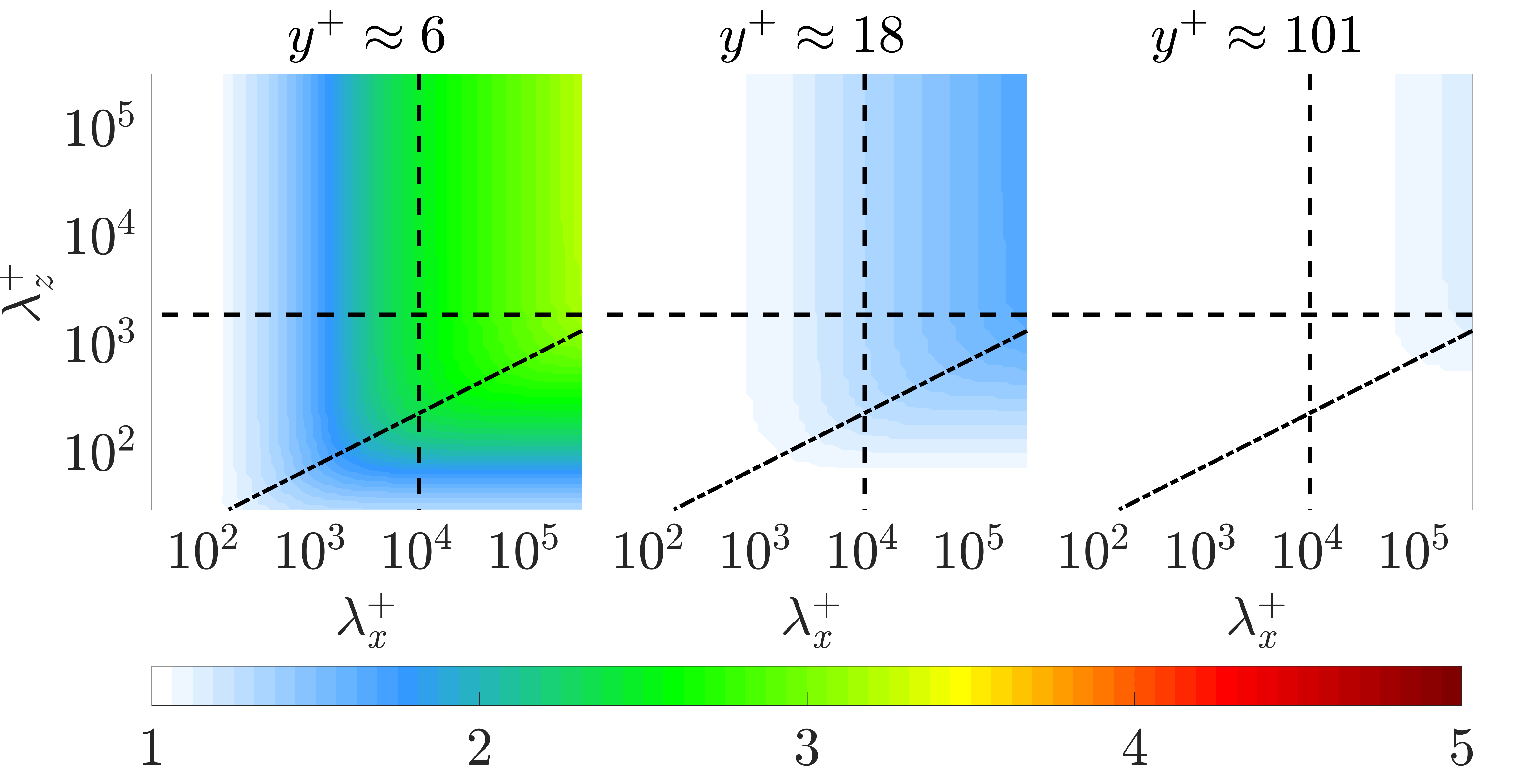}
		
\end{subfigure} 
\caption{ $\cliu{u}_c(y,\lambda_x,\lambda_z)/\bar{u}(y)$ \cliub{predicted using the viscous correction to Taylor's hypothesis in equation (\ref{viscousonly})} at (a) $\Rey_{\tau}=550$, (b) $\Rey_{\tau}=1000$, and
(c) $\Rey_{\tau}=5200$. 
The black dashed lines are given by $(\lambda_x,\lambda_z) = (2,0.4)$.
The black dash-dot lines are $\lambda_z^+ = \frac{5}{2}\sqrt{\lambda_x^+}$, which fits through the knee of these contours.}
	\label{Fig:convVisOnlyReCompare}
\end{figure}

The convective velocities computed with this viscous correction to Taylor's hypothesis for a range of Reynolds numbers are compared in figure \ref{Fig:convVisOnlyReCompare}. The results indicate that the regions in ($\lambda_x^+$, $\lambda_z^+$) where the convective velocities deviate from the local mean velocity are very similar across these Reynolds numbers, which is consistent with the observations in figure \ref{uc_nearWall_tripanel_localmean} indicating that the viscous correction preserves the previously observed Reynolds number invariance.

\cliuf{The viscous correction proposed here} may also be applicable to input-output based computations of convective velocities for other fluctuating quantities due to the similarity in the behavior of the near-wall convective velocities of velocity and vorticity components previously reported in the literature; see e.g., figures 3 and 5 of \citet{Geng2015}; figures 1 and 2 of \citet{Kim1993} and the results in \citet{liu2019vorticity}. Exploring this notion is a topic of ongoing work.

\section{Conclusion}
\label{sec:conclusion}

In this work, we \cliuf{analyze} convective velocities of fluctuating quantities based on \cliue{the} stochastically-forced linearized Navier--Stokes equations with \cliue{a given turbulent} mean velocity profile. \rev{This approach allows \cliuf{for} a detailed investigation of the scale-dependent convective velocities} at all wall-normal locations, which enables a comprehensive examination of the mechanisms at play in the generation of convective velocities.
 
The convective velocities of velocity fluctuations obtained using the input-output based model reproduce trends previously observed in the literature, such as the deviation of the average convective velocity from the mean velocity and its tendency toward a constant value in the near-wall region. The model-based results indicate that the convective velocity of the streamwise velocity fluctuations closer to the wall show a stronger dependence on wavelength.  The model predicted convective velocities show Reynolds number invariance when normalized in inner units, which is connected to the inner unit scaling of the resolvent operator \citep{Moarref2013} and consistent with observations from DNS data \citep{Geng2015} and experimental measurements \citep{Marusic2007}.

\rev{\cliue{Our analysis also indicates} that a wide range of structures contribute to the convective velocity especially in the viscous sublayer\rev{,} where the convective velocity has been shown to be strongly scale-dependent.}

The primary structures contributing to the near-wall convective velocity \cliue{based on the model} \rev{are larger} than the height of the buffer layer and are inclined at an angle between $25^\circ$ and $33^\circ$. These predictions confirm the findings of  \citet{Kim1993}\rev{,} who suggested that buffer layer structures are responsible for elevated convective velocities near the wall. However, \cliue{our analysis suggests} that a range of larger structures also contribute to \rev{this} near-wall convective velocity. \rev{We demonstrate that these} structures are self-similar in the cross-plane, similar to Townsend's attached-eddies, yet scale as the $\frac{2}{3}$ power of a cross-plane dimension in the streamwise direction. Our model suggests that there is a connection between the convective velocity and structures whose signatures in measurements of power spectra scale as $\lambda_z^+ \sim {\lambda_x^+}^\frac{2}{3}$.

We isolate and quantify the contributions from the pressure, mean shear, and viscous terms to the deviation of convective velocity from the \cliue{local mean velocity.} Based on this term- by-term analysis, a viscous correction to Taylor's hypothesis is proposed. The proposed correction leads to a simplified model that accurately reproduces the behavior of near-wall convective velocity of the streamwise velocity fluctuations of large-scale structures. 

The results presented here could be extended in a number of ways. For example, the representation of the forcing could be more closely tied to the nonlinearity observed in experimental \cliuf{or} numerical simulation results by e.g., \cliuf{using simulation data to  generate correlations for colored forcing \citep{Moarref2014,Zare2017}. Introducing an eddy viscosity based LNS representation \citep{reynolds1972mechanics} is another direction of ongoing work.} \rev{The present \rev{approach} \rev{has been specifically developed for wall-bounded flows with two homogenous spatial directions, and its efficacy has been demonstrated in the particular case of turbulent channel flow.} The applicability of such a model, and other stochastically-forced models based on the linearized Navier--Stokes equations to a broader class of turbulent flows, including turbulent boundary layers, \rev{is the subject of ongoing work.}
 }

\section*{Acknowledgments}
The authors thank Ismail Hameduddin \cliue{for his contributions to the formulation and preliminary version of these results. His insight through discussions are also greatly appreciated.} The authors gratefully acknowledge support from US National Science Foundation (NSF) through grant number CBET 1652244, program manager Ronald Joslin. C.L. also greatly appreciates support from the Chinese Scholarship Council.

\bibliographystyle{jfm}

\bibliography{main}

\end{document}